\documentclass[12pt]{IEEEtran}
\usepackage{cite}
\usepackage[dvips]{graphicx}
\usepackage[cmex10]{amsmath}
\usepackage{array}
\usepackage{subfig}
\usepackage{url}
\usepackage{amsfonts}
\usepackage{times}
\usepackage[usenames]{color}
\usepackage{verbatim}
\usepackage{ellipsis}
\usepackage{setspace}
\begin{document}
\title{Scalable Device-to-Device Communications For Frequency Reuse $>>$ 1}
\author{Daniel~Verenzuela and Guowang Miao
\thanks{D. Verenzuela is with the Division of Communication Systems in the Department of Electrical Engineering (ISY), Link\"{o}ping University (LiU), Link\"{o}ping, Sweden e-mail: daniel.verenzuela@liu.se.}
\thanks{G. Miao is with the Department
of Communication Systems, Royal Institute of Technology (KTH), Stockholm, Sweden e-mail: guowang@kth.se.}
\thanks{This project was supported by KTH ICT-TNG and was carried out as a thesis project of D. Verenzuela at KTH.}
}

\maketitle

\begin{abstract}

Proximity based applications are becoming fast growing markets suggesting that Device-to-Device (D2D) communications is becoming an essential part of future mobile data networks. We propose scalable admission and power control methods for D2D communications underlay cellular networks to increase the reuse of frequency resources and thus network capacity while maintaining QoS to all users. The aim of the proposed methods is to maximize the number of D2D links under QoS constraints, therefore maximizing network frequency reuse, while considering different levels of complexity and available channel state information (CSI) in a multi-cell environment. Numerical results show that by using D2D and the proposed multi-cell interference coordination and low power transmission method, the network spectral efficiency can be increased by as much as ten times, while low outage probability can be assured to provided QoS for all users.

\end{abstract}	

\begin{IEEEkeywords}
device-to-device communication, ICIC, multi-cell, admission control, transmit power control, spectral efficiency.
\end{IEEEkeywords}

%
\IEEEpeerreviewmaketitle

\section{Introduction}
\label{Introduction}

%
%
%
%

\IEEEPARstart{F}{r}equency reuse enables reusing the same frequency in cellular networks to support simultaneous communication demands on the same frequency and thus improves network capacity and spectral efficiency (SE). In the past decades, the global demand on mobile data traffic has increased considerably. The increase in network frequency reuse has been the driving force to boost network capacity to accommodate the traffic demand. It is expected that global mobile data traffic will continue its remarkable growth in the coming decades, most of which come from short distance and indoor communications \cite{cisco_2014}. Thus new techniques need to be developed to further increase network frequency reuse and the network capacity. Device-to-Device (D2D) communications has been proposed to improve the performance of cellular networks by allowing devices to communicate directly without relaying traffic to the base station (BS). This feature provides the means of increasing both spectral and energy efficiency, reducing delay, improving coverage and supporting new proximity based applications \cite{Design_D2D_journal}.  In addition, D2D enables a low cost solution for reusing frequency in short distance communications and thus is a good candidate technology to push the frequency reuse of cellular networks to far greater than one. Besides, from the service demand perspective, the emergence of new applications based on geographical proximity is becoming a fast growing market \cite{Market_ProSe} suggesting that D2D communications will become an important part of future mobile data networks. Thus studying the scalability of D2D communications is of paramount importance to accommodate future traffic demands.


Allowing D2D links to share the resources with cellular user equipments (CUEs) creates two levels of networks. The first level is the primary cellular network that is comprised of the communications between CUEs and their respective base stations (BSs). The secondary network level is composed of the simultaneous D2D links that share the resources of the primary cellular network. The main idea is to reutilize the resources of the primary cellular network as much as possible while minimizing the impact on its performance. For this reason we assume that D2D links will only share the uplink resources of the primary cellular network. Notice that when downlink resources are shared the D2D links may cause strong interference towards the CUEs whereas in the case of sharing uplink resources the interference caused by D2D links affects the BS, and this interference can be controlled by the BS \cite{Design_D2D_journal}. The sharing of resources between the two network levels allows for higher frequency reuse and SE. However as the reutilization of resources becomes higher the interference levels may increase to a point where the performance of both cellular and D2D networks is seriously degraded. Thus one of the main limitations on the scalability of D2D communications is the interference management.

When D2D links are added to the network, two main levels of interference are found:
\begin{enumerate}
\item Interference caused by the cellular network
	\begin{itemize}
	\item From CUEs towards other BSs (inter-cell interference).
	\item From CUEs towards D2D links.
	\end{itemize}
\item Interference caused by the D2D network
	\begin{itemize}
	\item From D2D links towards the BSs.
	\item From D2D links towards other D2D links.
	\end{itemize}
\end{enumerate}

In the first level we mainly have the inter-cell interference. In the uplink of the last generation cellular networks the resources within each cell are allocated orthogonally resulting in zero intra-cell interference. However the resources are shared by several cells causing inter-cell interference between the CUEs and BSs of different cells. This problem is well known and there has been important research done in the last years, e.g., in the case of the uplink of Long Term Evolution (LTE) networks the authors in \cite{impact_ICIC} analyze the impact of inter-cell interference coordination (ICIC) while considering the effects of power limitations, radio resource management functions, fast retransmissions, power control and adaptive modulation and coding methods. The work \cite{ASFR} proposes an adaptive soft frequency reuse method that decreases inter-cell interference improving the average throughput per user. In \cite{Coop_ICIC} the authors propose an interference aware joint scheduling method based on proportional fairness. The work \cite{ICoo_av} studies the problem of resource allocation considering the impact of inter-cell interference while maintaining a frequency reuse of one. The works \cite{OLFPC} and \cite{LTE_PC_VTC} perform an evaluation the LTE open loop fractional power control (OFPC) and the closed loop power control respectively considering the impact of inter-cell interference while giving an insight to the proper configuration of the design parameters.

The second effect of the first level interference refers to the impact of cellular communications on the D2D network. Since we assume the reuse of uplink resources, the CUEs may cause strong interference towards D2D links. However given that the cellular network is considered to have higher priority we do not consider modifying its functionality to accommodate the D2D network, rather the opposite. In addition, a D2D pair can always switch to the cellular mode if it is receiving too much interference from CUEs. Thus in this study we want to focus on the effects caused by the D2D network.

In the second level of interference we consider two main impacts. First the D2D links cause aggregated interference towards the BSs that may compromise the QoS of the uplink of CUEs. Secondly the D2D links may cause strong interference among each other limiting their QoS and reducing the SE of the network. Thus a careful design to deal with the second level interference is needed to obtain the most benefits of D2D communications and assure the QoS of both CUEs and D2D links.

To solve the problem of coordinating the interference, dynamic resource allocation algorithms have been presented in \cite{Capacity_D2D}, allowing an increase of capacity. Also joint resource block (RB) scheduling \cite{Res_allc_journal} and interference avoidance methods \cite{NEW_Int_cancellation} have been proposed to increase SE and reduce harmful interference respectively. Power control techniques have also been proposed to coordinate the interference, the work \cite{Comp_PC} conducts a comparative study of LTE power control techniques applied to D2D communications. A double threshold power control algorithm is proposed in \cite{Uplink_PC_D2D} to maximize the system throughput. The authors in \cite{Fuzzy_PC_D2D} proposed a continuous fuzzy logic power control method to limit the interference and improve the QoS of D2D links. In \cite{NEW_capacity_cooperation} the capacity of the system is studied under cooperative and non-cooperative interference coordination methods.

However in all of these works the number of D2D links that share the resources with the CUEs is always fixed or set. The network capacity can indeed be dramatically improved by allowing more D2D links in the network as it increases frequency reuse by reusing the same frequency more times for sending different data streams. The capacity in terms of the maximum number of D2D links that system can support has only been considered by few studies \cite{Res_all_max_D2D, DPC_D2D, Int_const_D2D}. In \cite{Res_all_max_D2D} a greedy heuristic resource allocation algorithm to maximize the number of D2D links in a single-cell is presented. The results show that by allowing more D2D links to share the same resources a considerable increase in SE can be achieved, however in this work power control is not considered and full channel state information (CSI) is assumed which may not be scalable in practical implementations. The study \cite{DPC_D2D} proposes and evaluates a series of distributed power control algorithms in D2D communications. However the implementation of the admission control procedures in terms of signaling is not mentioned and the impact on network scalability is not fully addressed.


In \cite{Int_const_D2D}, we have studied the feasibility of admitting several D2D pairs to share the resources of a CUE considering only the number of D2D links in a single-cell. Optimal and suboptimal solutions are proposed to coordinate the interference with the goal of maximizing the number of active D2D links in the system. The results show that allowing several D2D links to share the same recourses with CUEs without any limitation is indeed possible and it increases network SE.

It is worth mentioning that the studies \cite{Res_all_max_D2D, DPC_D2D, Int_const_D2D} are all focused on a single-cell scenario neglecting the impact of inter-cell interference. However in real applications this impact can be significant and needs to be considered. Another important remark mentioned in a comprehensive survey for D2D communications \cite{Survey_D2D} is that in most available literature the existing interference coordination solutions assume full CSI to be known at the BS. However this is usually not practical due to the signaling overhead, especially in multi-cell environments.

This study proposes admission control, inter-cell interference coordination, and power allocation methods for D2D communications underlay cellular networks to increase the network frequency reuse far beyond a factor of one by enabling as many simultaneous D2D links to communicate at the same time as possible. The proposed methods aim at maximizing the number of active D2D links subject to QoS constraints in a multi-cell environment of any size while considering three cases: no CSI, partial CSI or full CSI  available. In the first case we calculate a theoretical upper bound for the number of D2D links and implement a blind admission control (BAC) method considering average QoS requirements for CUE and D2D links. For the second case we present a distributed admission control (DAC) method to further improve the performance where each D2D pair decides its mode based on transmission power constraints to assure QoS for all users, the DAC methods is compatible with LTE standard and can be easily implemented in 5G LTE systems. In the last case we present an optimal admission control (OAC) method, which is based on exhaustive search and serves as the performance benchmark.

In our analysis we bound the number of D2D links based on approximate SINR expressions by considering the expected value of interfering terms. An alternative approach could be to use chance constrained programming to establish QoS constraints as outage probabilities \cite{chance_prog_Ad_C}. However, this problem formulation is not always tractable and may be expensive to solve in terms of computational resources. Our approach yields tractable and practical results that can be easily applied in current wireless networks. In particular, both BAC and DAC are based in algorithms that have extremely low signaling overhead and can be implemented in a multi-cell network regardless of its size. Therefore they constitute highly scalable solutions that can be easily implemented in the next generation networks, e.g., LTE Release 13 and onward.


The remainder  of this paper is presented as follows: section \ref{system_model} depicts the system model;In section \ref{Stat_Int_model} a statistical model for the interference is introduced; section \ref{N_CSIC} presents the analysis conducted to obtain the BAC method; section \ref{P_CSID} shows the analysis conducted to obtain DAC method; section \ref{F_CSIOp} presents the formulation of the OAC method; section \ref{Num_results} depicts the numerical results and analysis; finally section \ref{Conc} concludes our work.
%
\section{System Model}
\label{system_model}

Consider a multi-cell system where a number of D2D pairs are available in each cell. The D2D pairs and CUEs are randomly distributed in all cells. Our goal is to find the maximum number of D2D pairs that can communicate at the same time, thus maximizing the frequency reuse of the network. We focus on a single RB scenario where only one CUE is considered in each cell. We assume that resource allocation for the primary cellular network has already been implemented. We define $\phi_{xk} \in \{0,1\}$ $\big(\forall x\in\{ 1,\,...,\, N\},\,\forall k\in\{ 1,\,...,\,\hat{N}_{x}\}\big)$, as a binary random variable that indicates the state of each link $k$ in cell $x$. The parameter $N$ corresponds to the number of cells in the system and $\hat{N}_{x}$ is the number of available D2D links in cell $x$. When $\phi_{xk}=1$ the D2D link is active, otherwise $\phi_{xk}=0$. The frequency reuse factor is the rate at which the same frequency can be used in the network. Considering D2D, the frequency reuse factor of a cell $x$ is given by $K = \sum_k^{\hat{N}_x} \phi_{xk} + 1$. To maximize the network frequency reuse $K$, the objective is equivalent to maximizing $\sum_{x}^{N}\sum_k^{\hat{N}_x} \phi_{xk}$, that is, finding the maximum number of D2D links that can be in active communications and their corresponding transmission power levels while the QoS could be assured to all active users.

The maximum level of interference that can be tolerated in the system is given by the signal-to-interference plus noise ratio (SINR) constraints for the CUEs and D2D communications, depicted in \eqref{eq:CUE_SINR_c} and \eqref{eq:D2D_SINR_c} respectively. We also consider an upper bound for the transmission power of D2D links shown in \eqref{eq:D2D_max_P_c}.
\begin{subequations}
\label{eq:Int_const}
\begin{equation}
\Gamma_{x0}=\,\frac{P_{x0}G_{x0x0}}{I_{x0}^{D2D}\,+\,I_{x0}^{CUE}\,+\, \mathcal{N}_{BS}}\,\geq\,\gamma_{x0}^{th},
\label{eq:CUE_SINR_c}
\end{equation}
\begin{equation}
\Gamma_{xk}\,=\,\frac{\phi_{xk}P_{xk}G_{xkxk}}{I_{xk}^{D2D}+I_{xk}^{CUE}+\mathcal{N}_{D}}\,\geq\,\gamma_{xk}^{th},
\label{eq:D2D_SINR_c}
\end{equation}
\begin{equation}
\normalsize{\phi_{xk}P_{xk}\,\leq\,P_{D}^{max},
}
\label{eq:D2D_max_P_c}
\end{equation}
\end{subequations}
\begin{center}\normalsize{$\forall x\in\{ 1,\,...,\, N\}\;\forall k\in\{ 1,\,...,\,\hat{N}_{x}\}. $}\end{center}
The terms $I_{x0}^{D2D}$ and $I_{x0}^{CUE}$ correspond to the interference received at the BS of cell $x$, from the D2D links and CUEs respectively. Similarly $I_{xk}^{D2D}$ and $I_{xk}^{CUE}$ correspond to the interference received at the D2D link $k$ of cell $x$ from other D2D links and CUEs respectively. $\mathcal{N}_{BS}$ and $\mathcal{N}_{D}$ are the noise power at the BS and D2D links receivers respectively. $P_{x0}$ corresponds to the transmission power from the CUE at cell $x$, $P_{xk}$ is the power of the transmitting device of D2D pair $k$ in cell $x$ and $P_{D}^{max}$ is the maximum transmission power of D2D links. $\gamma_{x0}^{th}$ and $\gamma_{xk}^{th}$ represent the target SINR of the CUE uplink and the D2D link $k$ in cell $x$ respectively.

To describe the channel gains the following nomenclature is implemented: $G_{abij}$ corresponds to the path gain from the transmitter $b$ in cell $a$ to the receiver $j$ in cell $i$. Note that in all variables, CUE and BSs are indexed as $0$ and D2D users are indexed with non-negative integer numbers. In equations (\ref{eq:CUE_SINR_c}) and (\ref{eq:D2D_SINR_c}), $G_{x0x0}$ corresponds to the channel gain between the CUE and the BS of cell $x$, while $G_{xkxk}$ corresponds to the channel gain within the D2D pair $k$ in cell $x$, i.e., between the D2D transmitter and receiver of the same D2D pair. Thus we define the interference terms as:
\begin{subequations}
\label{eq:Int_terms}
\begin{equation}
\normalsize{
I_{x0}^{D2D}=\sum_{i=1}^{N}\sum_{j=1}^{\hat{N}_{i}}\phi_{ij}P_{ij}G_{ijx0},
}
\label{eq:I_x0_D2D_c}
\end{equation}
\begin{equation}
\normalsize{
I_{x0}^{CUE}=\sum_{\substack{i=1\\i\neq x}}^{N}P_{i0}G_{i0x0},
}
\label{eq:I_x0_CUE_c}
\end{equation}
\begin{equation}
\normalsize{
I_{xk}^{D2D}=\sum_{i=1}^{N}\sum_{j=1}^{\hat{N}_{i}}\phi_{ij}P_{ij}G_{ijxk} - \phi_{xk}P_{xk}G_{xkxk},
}
\label{eq:I_xk_D2D_c}
\end{equation}
and
\begin{equation}
\normalsize{
I_{xk}^{CUE}=\sum_{i=1}^{N}P_{i0}G_{i0xk},
}
\label{eq:I_xk_CUE_c}
\end{equation}
\end{subequations}
\begin{center}\normalsize{$\forall x\in\{ 1,\,...,\, N\}\;\forall k\in\{ 1,\,...,\,\hat{N}_{x}\}. $}\end{center}

Here we can see that the terms $I_{x0}^{D2D}$ and $I_{xk}^{D2D}$ correspond to the second level of interference related to the impact of the D2D network, where $I_{x0}^{D2D}$ represents the interference caused by D2D links to BSs and $I_{xk}^{D2D}$ the interference among different D2D links. These are the main effects that we want to control in order to provide scalable interference coordination methods for D2D communications.

On the other hand the terms $I_{x0}^{CUE}$ and $I_{xk}^{CUE}$ correspond to the first level of interference that is related to the impact of the cellular network, where $I_{x0}^{CUE}$  depicts the inter-cell interference and $I_{xk}^{CUE}$ the interference caused by CUEs towards D2D links. Since we are not interested in modifying the functionality of the cellular networks these terms are not part of the design variables for the interference coordination methods. Thus we redefine the SINR target for CUEs as:

\begin{equation}
\gamma_{x0}^{th}=\frac{\Gamma_{x0}^{i}}{\delta}=\frac{P_{x0}G_{x0x0}}{\left(I_{x0}^{CUE}+\mathcal{N}_{BS}\right)\delta},\; \forall \delta \in \{\mathbb{R}^{+}; \delta > 1\},
\label{eq:delta_CUE}
\end{equation}
where $\Gamma_{x0}^{i}$ is the SINR of CUEs before D2D links are added to the system. The parameter $\delta$ corresponds to the desired ratio between the CUE's SINR before and after D2D links are added, i.e., the SINR loss of CUEs due to the interference caused by D2D links. This definition allows a clear evaluation of the impact of D2D links to the CUEs uplink. The parameter $\delta$ can be broadcasted by the BS or sent to a D2D link when it is newly established. In our analysis we assume a fixed target SINR for all D2D links defined as $\gamma_{D}$ in order to implement fairness in the admission control of D2D links.

\section{Statistical Interference Model}
\label{Stat_Int_model}

To develop the interference coordination methods we assume different levels of CSI to be available. For the low complexity solutions, we need to establish statistical models for the interference and channel gains in order to account for the unavailable CSI.


\subsection{Interference Model}
\label{Int_model}

\begin{figure}[t]
\begin{center}
\includegraphics[width=0.4\textwidth]{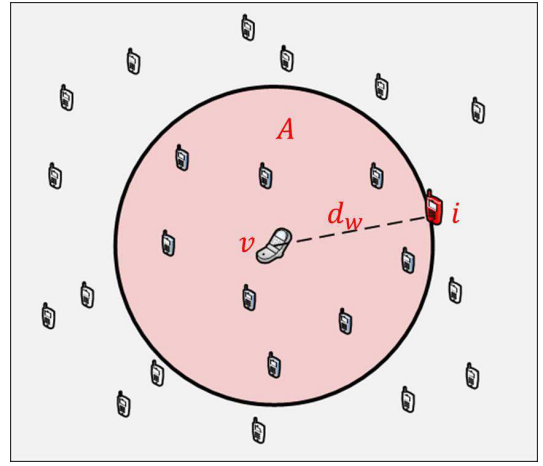}
\end{center}
\caption{Interference Model.}
\label{fig:Int_area}
\end{figure}

Consider a victim receiver $v$ surrounded by $\tilde{N}$ devices, we define the aggregated interference received at $v$ as:
\begin{equation}
\normalsize{
I_{v}\,=\,\sum_{i=1}^{\tilde{N}}P_{tx_{vi}}G_{vi},
\label{eq:I_v}
}
\end{equation}
where $P_{tx_{vi}}$ is the transmission power of an interfering device $i$ and $G_{vi}$ is the channel gain between $v$ and $i$. To find a statistical model for the interference we assume that the interfering devices are randomly distributed within a given area $A$, as shown in Fig \ref{fig:Int_area}. Thus the channel gains can be represented as a random variable $G_{vi}$. We also assume that the interfering devices have the same transmission power $ P_{tx_{vi}} = P_{tx} \leq P_{max}$, where $P_{max}$ is the maximum transmission power allowed for the devices.\footnote{Note that this assumption is only made for the estimation of the aggregated interference. In the proposed interference coordination methods power control is implemented.} Thus we can define an expected value for the aggregated interference within $A$ as:
\begin{equation}
\normalsize{
\mathbb{E}[I_{v}]\,=\,\tilde{N}_{A}\, A\, P_{tx}\,\mathbb{E}[G_{vi}],
\label{eq:I_v_2}
}
\end{equation}
where $\tilde{N}_{A}$ is the density of interfering devices per unit area.

In order to obtain reasonable values for $\tilde{N}_{A}$ the area $A$ needs to be finite, which means that in practical applications there will be interfering devices outside of $A$, as shown in Fig \ref{fig:Int_area}. We define $A = \pi (d_{w})^{2}$ as a circular interference area around $v$ where $d_{w}$ is the maximum distance between $v$ and an interfering device. Notice that the interference caused by the devices outside of $A$ is negligible compared to the one caused by the users inside due to the path loss attenuation.

 To determine the value of $d_{w}$ we assume that if the received power from an interfering device is lower than a threshold, then its effect can be neglected. Notice that the received power of the interfering devices depends also on their transmission power and this is meant to be  used as a design variable in the later analysis. Thus we calculate $d_{w}$ considering the maximum transmission power that is allowed $P_{max}$ so that the interference area $A$ is obtained for the worst interference scenario. Then we consider an interferer $w$ that is located at the edge of the interference area and we define $d_{w}$ as:
\begin{equation}
\normalsize{
\begin{array}{c}
P_{max}\mathbb{E}[G_{vw}]\leq \mathcal{N}_{v},\\[2mm]
d_{vw}\geq\text{\Large{$\left(\frac{P_{max}c_{v}\mathbb{E}[|h_{vw}|^{2}]}{\mathcal{N}_{v}}\right)^{1/\alpha_{v}} $}}=d_{w},
\end{array}
}
\label{eq:d_w}
\end{equation}
where $\mathcal{N}_{v}$ is the noise power at the victim receiver and $d_{vw}$ is the distance between devices $v$ and $w$. We define the channel gain between $v$ and $w$ as:
\begin{equation}
\normalsize{
G_{vw} = c_{v}d_{vw}^{-\alpha_{v}}|h_{vw}|^{2},
}
\label{eq:Channel_gain}
\end{equation}
where $c_{v}$ refers to a propagation constant and $\alpha_{v}$ is the path loss exponent. The effects of fast fading are represented by $|h_{vw}|^{2}$.
\subsection{Channel Gain Model}
\label{channel_gain_model}

If CSI is not available we can model the channel gain between two devices $v$ and $i$ as a random variable $G_{vi}$, defined in \eqref{eq:Channel_gain}, where $d_{vi}$ and $h_{vi}$ are independent random variables. Thus we can calculate the expected value of $G_{vi}$ as:
\begin{equation}
\normalsize{
 \mathbb{E}[G_{vi}] = c_{v} \mathbb{E}[d_{vi}^{-\alpha_{v}}] \mathbb{E}[|h_{vi}|^{2}].
}
\label{eq:E_G_D2D}
\end{equation}
Notice that in our analysis we want to establish statistical models of the channel gains. So consider the channel to be invariant during the period of interest, thus we assume $\mathbb{E}[|h_{vi}|^{2}] = 1$.

We also assume device $v$ to be located at a fixed point and device $i$ to be positioned randomly following a circular distribution around $v$. Thus the probability density function  of $d_{vi}$ is given by a triangular distribution depicted as:
\begin{equation}
\normalsize{
 f_{d_{vi}}(x)= \Bigg\{\begin{array}{cl}\frac{\text{\normalsize{$2x$}}}{\text{\normalsize{$(d_{max})^{2}$}}} & \text{if}\quad d_{min}\leq x\leq d_{max}, \\[3mm] 0 &\text{otherwise}.\end{array}
}
\label{eq:pdf_d}
\end{equation}
Combining \eqref{eq:E_G_D2D} and \eqref{eq:pdf_d} we have that $\forall \alpha_{v} \in\{\mathbb{R}^{+}; \alpha_{v} > 2\} $.
\begin{equation}
\normalsize{
\begin{array}{l}
\mathbb{E}[G_{vi}] =  c_{v} \text{\large{$\int_{\!d_{min}}^{d_{max}}x^{-\alpha_{v}} f_{d_{vi}}$}}(x) dx\\[5mm]
\qquad\quad=c_{v}\text{\large{$\int_{\!d_{min}}^{d_{max}}\frac{2 x^{(1-\alpha_{v})}}{(d_{max})^{2}}$}}d x\\[5mm]
\qquad\quad=\text{\large{$\frac{2c_{v}\left(d_{min}^{-(\alpha_{v}-2)}-\:d_{max}^{-(\alpha_{v}-2)}\right)}{(d_{max})^{2} \left(\alpha_{v}-2\right)}$}}.
 \end{array}
}
\label{eq:E_G_D2D_r}
\end{equation}
This result is applied to model all channel gains considered in this investigation. Note also that for practical applications the probability density function of $d_{vi}$ can be changed to match real user distributions.




\section{Blind Admission control}
\label{N_CSIC}

In practical applications obtaining CSI is not always possible because of high signaling overhead. Particularly if we consider the case of D2D communications, having CSI from every D2D link in the system would considerably increase the signaling overhead. Thus we present the BAC method where no CSI is necessary. Admission control here refers to letting a pair of devices in proximity communicate in the D2D mode.

In the BAC method each BS independently estimates an upper bound for the number of D2D links that can be active in its cell based on average constraints for the QoS of CUEs and D2D links. Then the active D2D links are selected randomly from the available ones within the cell. The transmission power of D2D links is obtained by applying the channel inversion power control algorithm which allows for a fixed received power at the receiving device. Thus the transmission power of a given D2D link $k$ in a cell $x$ is depicted as:
\begin{equation}
P_{xk} = \frac{P_{r_{D}}}{G_{xkxk}}\,\leq\,P_{D}^{max},\begin{array}{c}
\forall x\in\{ 1,\,...,\, N\}\\
\forall k\in\{ 1,\,...,\,\tilde{N}_{x}\}
\end{array},
\label{eq:P_rx_D2D}
\end{equation}
where $P_{r_{D}}$ is the received power for all D2D links which is calculated and broadcasted by the BS. Notice that the channel gain between devices of the same pair $G_{xkxk}$ is known to them but unknown to the BS.

To obtain an upper bound for the number of D2D links first we assume $\phi_{xk} = 1\forall x,k$ and calculate the expected value of interference constrains \eqref{eq:CUE_SINR_c} and  \eqref{eq:D2D_SINR_c} combined with \eqref{eq:delta_CUE} and \eqref{eq:P_rx_D2D}. As a result we have:
\begin{subequations}
\label{eq:E_Int_const}
\begin{equation}
\frac{P_{x0}G_{x0x0}}{\mathbb{E}[I_{x0}^{D2D}]\,+\,\mathbb{E}[I_{x0}^{CUE}]\,+\, \mathcal{N}_{BS}}\,\geq\,\frac{\Gamma_{x0}^{i}}{\delta},
\label{eq:E_CUE_SINR_c}
\end{equation}
\begin{equation}
\frac{P_{r_{D}}}{\mathbb{E}[I_{xk}^{D2D}]+\mathbb{E}[I_{xk}^{CUE}]+\mathcal{N}_{D}}\,\geq\,\gamma_{D},
\label{eq:E_D2D_SINR_c}
\end{equation}
\end{subequations}
\begin{center}\normalsize{$\forall x\in\{ 1,\,...,\, N\}\;\forall k\in\{ 1,\,...,\,\hat{N}_{x}\}. $}\end{center}
Since no CSI is available we consider the channel gains to be random variables, then by applying the interference model described in section \ref{Int_model} we can define the interference terms that depend on D2D links as:
\begin{equation}
\mathbb{E}[I_{x0}^{D2D}]\,=\,\tilde{N}_{A}\, A_{C}\,P_{r_{D}}\,\frac{\mathbb{E}[G_{D2D-BS}]}{\mathbb{E}[G_{D2D}]},
\label{eq:E_I_x0_D2D_c}
\end{equation}
\begin{equation}
\begin{array}{c}
\mathbb{E}[I_{xk}^{D2D}]\,=\,\left(\tilde{N}_{A}\, A_{D} - 1\right)\, P_{r_{D}}\,\frac{\mathbb{E}[G_{D2D-I}]}{\mathbb{E}[G_{D2D}]},\\
\forall x\in\{ 1,\,...,\, N\}\;\forall k\in\{ 1,\,...,\,\hat{N}_{x}\}.
\end{array}
\label{eq:E_I_xk_D2D_c}
\end{equation}

%

%
The term $G_{D2D}$ is a random variable that represents the channel gain between two devices of the same D2D pair. Similarly $G_{D2D-BS}$ describes the channel gain between D2D transmitters and the BS. $G_{D2D-I}$ represents the channel gain between D2D transmitters and a given D2D receiver from different D2D pairs. $A_{D}$ and $A_{C}$ are the interference areas for D2D links and CUEs constraints respectively. $\tilde{N}_{A}$ represents the density of active D2D links per unit area. Note that in \eqref{eq:E_I_xk_D2D_c} the number of D2D links is subtracted by one because there needs to be more than one D2D link in $A_{D}$ to cause interference.

Applying \eqref{eq:E_I_x0_D2D_c} and \eqref{eq:E_I_xk_D2D_c} to \eqref{eq:E_CUE_SINR_c} and \eqref{eq:E_D2D_SINR_c} allows us to obtain two statistical upper bounds of the density of D2D links per unit area that can be accepted in the system. One is obtained from the CUE constraint $\tilde{N}_{C}^{UB}$ and another from the D2D constraint $\tilde{N}_{D}^{UB}$, depicted as:

\begin{subequations}
\label{eq:E_UB}
\begin{equation}
\tilde{N}_{A}\leq\frac{\mathbb{E}[G_{D2D}]I_{C}}{P_{r_{D}}A_{C}\mathbb{E}[G_{D2D-BS}]}=\tilde{N}_{C}^{UB},
\label{eq:N_A_CUE}
\end{equation}
\begin{equation}
\normalsize{
\text{\small{$\tilde{N}_{A}\leq\left(\frac{\mathbb{E}[G_{D2D}]}{\mathbb{E}[G_{D2D-I}] \gamma_{D}}+ 1 - \frac{\mathbb{E}[G_{D2D}]I_{D}}{P_{r_{D}}\mathbb{E}[G_{D2D-I}]}\right) \frac{1}{A_{D}}=\tilde{N}_{D}^{UB}$}},
}
\label{eq:N_A_D2D}
\end{equation}
where
\begin{equation}
\normalsize{
I_{C} = \left(\delta - 1\right)\left(\mathbb{E}[I_{x0}^{CUE}] + \mathcal{N}_{BS}\right) ,
}
\label{eq:I_C}
\end{equation}
\begin{equation}
\normalsize{
I_{D} = \left(\mathbb{E}[I_{xk}^{CUE}] + \mathcal{N}_{D}\right),
}
\label{eq:I_D}
\end{equation}
\end{subequations}
\begin{center}\normalsize{$\forall x\in\{ 1,\,...,\, N\}\;\forall k\in\{ 1,\,...,\,\hat{N}_{x}\}. $}\end{center}

\begin{figure}[t]
\begin{center}
\includegraphics[width=0.48\textwidth]{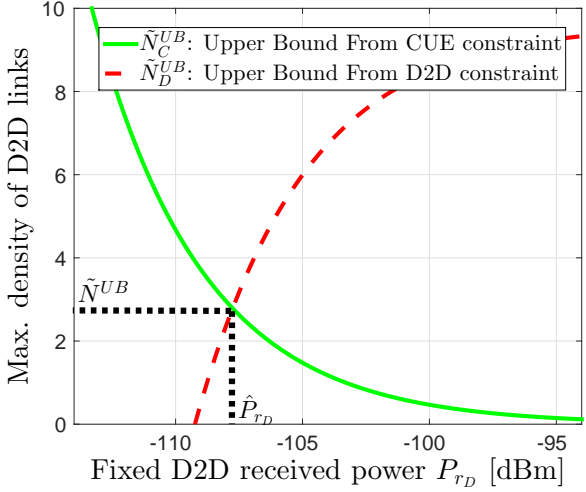}
\end{center}
\caption{Max. density of D2D links vs received power $P_{r_{D}}$ for $\delta = 3$[dB], $\gamma_{D}=10$[dB]. In a circular area of radius $R=400$ [m] and with $d_{D2D}$ between 10 [m] and 50 [m].}
\label{fig:N_UBvsP_D2D}
\end{figure}
These upper bounds depend on the received power of D2D links $P_{r_{D}}$, Fig \ref{fig:N_UBvsP_D2D} depicts a numerical example of $\tilde{N}_{C}^{UB} A$ and $\tilde{N}_{C}^{UB} A$ as the upper bounds for the number of D2D links in a circular area $A$ of radius $R$. The term $d_{D2D}$ is a random variable that represents the distance between the transmitter and receiver of a given D2D pair. $\tilde{N}_{C}^{UB}$ is monotonically decreasing with respect to $P_{r_{D}}$. $\tilde{N}_{D}^{UB}$ increases until it reaches a saturation point, which means that after certain value of $P_{r{D}}$, the density of D2D links per unit area cannot be increased for a D2D target SINR $\gamma_{D}$.

The overall upper bound for the density of D2D links is given by the minimum between $\tilde{N}_{C}^{UB}$ and $\tilde{N}_{D}^{UB}$ in order to satisfy both QoS constraints \eqref{eq:E_CUE_SINR_c} and \eqref{eq:E_D2D_SINR_c}. Thus it is possible to find $\hat{P}_{r_{D}}$ where $\tilde{N}^{UB} =\tilde{N}_{C}^{UB}=\tilde{N}_{D}^{UB}$ so that the density of D2D links is maximized. Solving for $\tilde{N}^{UB}$, the upper bound for the density of D2D links per unit area is:
\begin{equation}
\normalsize{
\tilde{N}^{UB}= \frac{I_{C}\left(\frac{\mathbb{E}[G_{D2D}]}{\gamma_{D}} + \mathbb{E}[G_{D2D-I}]\right)}{A_{D}\mathbb{E}[G_{D2D-I}]I_{C} + A_{C}\mathbb{E}[G_{D2D-BS}]I_{D}}
}
\label{eq:N_UB_a}
\end{equation}
where
\begin{equation}
\normalsize{
\hat{P}_{r_{D}}=\frac{\mathbb{E}[G_{D2D}]\left(A_{D}\mathbb{E}[G_{D2D-I}]I_{C} + A_{C}\mathbb{E}[G_{D2D-BS}]I_{D}\right)}{A_{C}\mathbb{E}[G_{D2D-BS}]\left(\frac{\mathbb{E}[G_{D2D}]}{\gamma_{D}} + \mathbb{E}[G_{D2D-I}]\right)}.
}
\label{eq:P_r_D2D_th}
\end{equation}
\begin{center}\normalsize{$\forall x\in\{ 1,\,...,\, N\}\;\forall k\in\{ 1,\,...,\,\hat{N}_{x}\}. $}\end{center}
It is worth mentioning that as shown in \eqref{eq:I_C} and \eqref{eq:I_D},  $I_C$ and $I_D$ depend on $\mathbb{E}[I_{x0}^{CUE}]$ and $\mathbb{E}[I_{xk}^{CUE}]$ respectively. These terms can be calculated by applying the interference model described in section \ref{Int_model}. In this case the density of users per unit area is known and corresponds to $1/A_{cl}$, where $A_{cl}$ is the area of the cell, given that there is only one CUE per cell.

From equation \eqref{eq:N_UB_a} we can see that the expected values of the channel gains determine the maximum density of D2D links per unit area that can be allowed in the system. By implementing the model found in section \ref{channel_gain_model} we define these terms as:
\begin{subequations}
\begin{equation}
\normalsize{
\mathbb{E}[G_{D2D}]=\frac{2c_{d}\left(d_{D2D_{min}}^{-(\alpha_{d}-2)}-\:d_{D2D_{max}}^{-(\alpha_{d}-2)}\right)}{(d_{D2D_{max}})^{2} \left(\alpha_{d}-2\right)},
}
\label{eq:Cen_G_D2D}
\end{equation}
\begin{equation}
\normalsize{
\mathbb{E}[G_{D2D-BS}]=\frac{2c_{0}\left(d_{D2D-BS_{min}}^{-(\alpha_{0}-2)}-\:d_{D2D-BS_{max}}^{-(\alpha_{0}-2)}\right)}{(d_{D2D-BS_{max}})^{2} \left(\alpha_{0}-2\right)},
}
\label{eq:Cen_G_D2D-BS}
\end{equation}
\begin{equation}
\normalsize{
\mathbb{E}[G_{D2D-I}]=\frac{2c_{d}\left(d_{D2D-I_{min}}^{-(\alpha_{d}-2)}-\:d_{D2D-I_{max}}^{-(\alpha_{d}-2)}\right)}{(d_{D2D-I_{max}})^{2} \left(\alpha_{d}-2\right)},
}
\label{eq:Cen_G_D2D-I}
\end{equation}
\end{subequations}
where  the distance between the D2D transmitter and receiver of the same pair is a random variable within $[d_{D2D_{min}}, d_{D2D_{max}}]$. The distance between the D2D links and the BS is randomly distributed in the interval $[d_{D2D-BS_{min}}, d_{D2D-BS_{max}}]$. The distance between D2D transmitters and receivers of different D2D pairs is randomly distributed within $[d_{D2D-I_{min}}, d_{D2D-I_{max}}]$. The term $\alpha_{0}$ corresponds to the path loss exponent for the channel between devices and the BS, whereas $\alpha_{d}$ corresponds to the path loss exponent for the channel between devices. The term $c_{0}$ refers to a propagation constant for the channel between devices and the BS, and $c_{d}$ corresponds to a propagation constant for the channel between devices.

Notice that the maximum limit for the distribution of the distances $d_{D2D-BS_{max}}$ and $d_{D2D-I_{max}}$ are given by the definition of the interference area (see section \ref{Int_model}, eq: \ref{eq:d_w}). In practical applications more sophisticated spatial distributions of users can be obtained in order to have more accurate values for the expectations of the channel gains.

At this point we are able to estimate the maximum number of D2D links that can be allowed in the system without considering CSI. In order to implement this result each BS estimates independently the number of D2D links that can be active in its cells as:
\begin{equation}
\normalsize{
\tilde{N}_{x}=\min \{\lfloor \tilde{N}^{UB}A_{cl_{x}} \rfloor, \,\hat{N}_{x} \}, \forall x\in\{ 1,\,...,\, N\}.
}
\label{eq:Cen_N_D2D}
\end{equation}

The term $A_{cl_{x}}$ is the area of cell $x$ and $\hat{N}_{x}$ is the number of available D2D links in the cell. Once the BS calculates the number of active D2D links $\tilde{N}_{x}$, it simply selects them randomly from the available ones and broadcasts the received power parameter $\hat{P}_{r_{D}}$ for the power control of D2D links, which can be done in the admission control of D2D communications.




  \section{Distributed Admission Control}
\label{P_CSID}

In the implementation of D2D communications there is a certain amount of CSI that is already available in the system. Thus we present the distributed admission control (DAC) method that makes use of the available information to better coordinate the interference between D2D links and CUEs.

The DAC method is based on a distributed algorithm where the D2D pairs decide independently their active status and their transmission power by adding a limited amount of semi-static signaling overhead. The main idea is that every D2D pair decides their own active status and transmission power based on general information parameters that are broadcasted by the BSs, e.g, number of active D2D links in the cells, path loss towards the BS, area of the cells, etc. In contrast with the BAC method where the admission control is done at random, this solution is able to adapt the admission and transmission power of each D2D link to better coordinate the interference.

In order to implement this method we use the same constraints defined in previous sections to derive an upper and lower bound for the transmission power of D2D links based on the QoS of CUEs and D2D links. Then each D2D link decides its active status depending on the feasibility of its transmission power constraints, i.e., being able to assure QoS for itself while maintaining the aggregated interference to the CUEs below a threshold.

To calculate the constraints for the transmission power of D2D links, we need a statistical estimation of the interference scenario given that CSI is limited. Thus we make use of the interference model found in section \ref{Int_model} which can be applied at the D2D pairs if the BSs broadcast the numbers of active D2D links in their cells.

In D2D communications underlay cellular networks the BS plays a role in the discovery procedure, hence we can assume that the active status of each D2D pair can be known to its serving BS. Thus the BS would have information about the number of active D2D links in their respective cells.

To illustrate the implementation of the DAC method, let us assume a D2D pair $k$ in a cell $x$, denoted by $D2D_{xk}$, that needs to decide its active status. Since each D2D link makes an independent decision with limited CSI, $D2D_{xk}$ assumes the same transmission power for all D2D links $P_{ij}=P_{D_{xk}},\:\forall i\in\{ 1,\,...,\, N\}, \forall j\in\{ 1,\,...,\,\hat{N}_{i}\}$.\footnote{Note that this assumption is only made to estimate the interference that each D2D link would received from other D2D links. However, in the implementation of the DAC method the transmission power of each D2D pair is different and decided in a distributed manner (see \eqref{eq:Dist_P_LB} and \eqref{eq:Dist_P_xk}).}

Initially we define an upper and lower bound for the transmission power of a D2D pair as $P_{D_{xk}}^{UB}$ and $P_{D_{xk}}^{LB}$ respectively. Then we compare the two sets $[-\infty, P_{D_{xk}}^{UB}]$ and $[P_{D_{xk}}^{LB}, \infty]$. If their intersection is a non-empty set, $D2D_{xk}$ is active $\phi_{xk} = 1$, otherwise $\phi_{xk} = 0$. This rule allows $D2D_{xk}$ to evaluate the feasibility of its link given that the upper bound limits the interference to the CUE uplink and the lower bound assures the QoS of $D2D_{xk}$ link. Notice that our objective is to maximize the number of active D2D links while assuring QoS to all users, thus $D2D_{xk}$ should only be in active mode if the two power sets intersect.

To obtain the upper bound first we redefine the term $I_{x0}^{D2D}$, found in \eqref{eq:I_x0_D2D_c},  as:
\begin{equation}
\normalsize{
I_{x0}^{D2D} = \phi_{xk}P_{xk}G_{xkx0}+\hat{I}_{x0}^{D2D} = \phi_{xk}P_{D_{xk}}G_{xkx0}+\hat{I}_{x0}^{D2D},
}
\label{eq:Dist_I_x0_D2D_c}
\end{equation}
where $\hat{I}_{x0}^{D2D}$ corresponds to the aggregated interference caused by active D2D links to the BS of cell $x$ ($BS_{x}$). Since $D2D_{xk}$ does not have CSI to calculate $\hat{I}_{x0}^{D2D}$, we consider it to be a random variable. Thus we can calculate its expected value by applying the model found on section \ref{Int_model}. As a result we have:
\begin{equation}
\normalsize{
\mathbb{E}[\hat{I}_{x0}^{D2D}]=\frac{\tilde{N}_{x}}{A_{cl_{x}}}\, A_{x0}\,P_{D_{xk}}\,\mathbb{E}[G_{D2D-BS}],
}
\label{eq:Dist_I_x0_D2D}
\end{equation}
where $A_{x0}$ is the interference area and $A_{cl_{x}}$ is the area of cell $x$. The term $\mathbb{E}[G_{D2D-BS}]$ is the expected value of the channel gain between active D2D links and BS. 

Finally we obtain a statistical upper bound for the transmission power of D2D links $P_{D_{xk}}^{UB}$ by combining the expected value of \eqref{eq:CUE_SINR_c} and \eqref{eq:D2D_max_P_c} with \eqref{eq:Dist_I_x0_D2D}, thus $P_{D_{xk}}^{UB}$ is defined as:
\begin{subequations}
\begin{equation}
\normalsize{
P_{D_{xk}}^{UB}=\min\left\{ \frac{\text{\normalsize{$\hat{I}_{x0}^{th}$}}}{\text{\normalsize{$G_{xkx0} + \frac{\tilde{N}_{x}}{A_{cl_{x}}} A_{x0} \mathbb{E}[G_{D2D-BS}]$}}} ,P_{D}^{max} \right\},
}
\label{eq:Dist_P_UB}
\end{equation}
\begin{equation}
\normalsize{
\hat{I}_{x0}^{th} = \left(I_{x0}^{th}-\mathbb{E}[I_{x0}^{CUE}]-\mathcal{N}_{BS}\right),
}
\label{eq:hat_I_th_x0}
\end{equation}
\end{subequations}
\begin{center}\normalsize{$\forall x\in\{ 1,\,...,\, N\}\;\forall k\in\{ 1,\,...,\,\hat{N}_{x}\}. $}\end{center}
where $G_{xkx0}$ corresponds to the channel gain between $D2D_{xk}$ and $BS_{x}$, which can be obtained by monitoring the downlink reference signals. The term $I_{x0}^{CUE}$ is considered to be a random variable and can be estimated by applying the interference model of section \ref{Int_model}.

The parameter $\hat{I}_{x0}^{th}$ is the amount of interference that the D2D links can cause to the $BS_{x}$ so that the QoS of the CUE can be assured. The term $I_{x0}^{th}$ is given by the definition of the CUE QoS based on the SINR loss, thus we have:
\begin{equation}
\normalsize{
I_{x0}^{th}=\frac{P_{x0}G_{x0x0}}{\gamma_{x0}^{th}}=\frac{\delta P_{x0}G_{x0x0}}{\Gamma_{x0}^{i}},\; \forall x\in\{ 1,\,...,\, N\},
}
\label{eq:Dist_I_x0_tha}
\end{equation}
however this information is not available at the D2D links in normal conditions, thus we assume it is broadcasted by $BS_{x}$. 

Note that $P_{D_{xk}}^{UB}$ can be easily implemented in LTE-A systems as the existing power control for D2D communications in LTE-A ensures the interference from D2D communications to the serving BSs to be at fixed tolerable levels, which are configured by the BSs  \cite{3gpp_36213}. The BSs can configure the parameters in the power control formula so that the power control value is the one in \eqref{eq:Dist_P_UB} to serve as the upper bound.

\begin{figure}[t]
\begin{center}
\hspace*{45pt}
\includegraphics[width=0.4\textwidth]{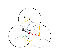}
\end{center}
\caption{Information obtained from three closest sectors of pair $D2D_{xk}$.}
\label{fig:sec_info}
\end{figure}

To obtain the lower bound for the transmission power we consider the constraint \eqref{eq:D2D_SINR_c}, where the term $I_{xk}^{D2D}$ represents the interference from active D2D links to $D2D_{xk}$. Similarly to the previous  analysis we can estimate this term as:
\begin{equation}
\normalsize{
\mathbb{E}[I_{xk}^{D2D}]=\frac{\tilde{N}_{xd}}{A_{dk}}\, A_{xk}\, P_{D_{xk}}\,\mathbb{E}[G_{D2D-I}].
}
\label{eq:Dist_I_xk_D2D}
\end{equation}
Here the parameter $A_{xk}$ is the interference area and $\mathbb{E}[G_{D2D-I}]$ is the expected value of the channel gain between an interfering D2D link (within $A_{xk}$) and $D2D_{xk}$. To estimate the number of active D2D links per unit area in the surroundings of $D2D_{xk}$ we assume that the cells can be divided into three sectors, which is highly common in practical applications. Thus $BS_{x}$ can know the number of active D2D links in each sector and this could be broadcasted to the users. $\tilde{N}_{xd}$ represents the sum of active D2D links in the three sectors that are closer to $D2D_{xk}$ and $A_{dk}$ is the area enclosed by such sectors, as illustrated in Fig. \ref{fig:sec_info}, a simple way of estimating $\tilde{N}_{xd}$ and $A_{dk}$ is:
\begin{equation}
\tilde{N}_{xd} = \tilde{N}_{xc} + \tilde{N}_{ya} +\tilde{N}_{zb},
\label{eq:DAC_N_sec}
\end{equation}
\begin{equation}
A_{dk} = A_{xc} + A_{ya} + A_{zb}.
\label{eq:DAC_A_sec}
\end{equation}
More advanced estimators can be used to obtain more accurate results.

By calculating the expected value of \eqref{eq:D2D_SINR_c} and combining it with \eqref{eq:Dist_I_xk_D2D} we can obtain an statistical lower bound for the transmission power of D2D links as:
\begin{equation}
\normalsize{
P_{D_{xk}}^{LB}=\frac{\text{\normalsize{$\left(\mathbb{E}[I_{xk}^{CUE}]+\mathcal{N}_{D}\right) \gamma_{D}$}}}{\text{\normalsize{$G_{xkxk} - \left(\gamma_{D}\frac{\tilde{N}_{xd}}{A_{dk}}A_{xk}\mathbb{E}[G_{D2D-I}]\right)$}}},
}
\label{eq:Dist_P_LB}
\end{equation}
\begin{center}\normalsize{$\forall x\in\{ 1,\,...,\, N\}, \forall k\in\{ 1,\,...,\,\hat{N}_{x}\}.$}\end{center}
The term $G_{xkxk}$ corresponds to the channel gain between the transmitter and receiver of $D2D_{xk}$ which is obtained from the discovery procedure.  The parameter $I_{xk}^{CUE}$ corresponds to the interference caused by CUEs towards $D2D_{xk}$ and can be estimated by applying the interference model of section \ref{Int_model}.

Finally the decision of $D2D_{xk}$ to be active is given by:
\begin{subequations}
\begin{equation}
\normalsize{
\phi_{xk}=\Bigg \{ \begin{array}{l} 1\; \text{if}\;  P_{D_{xk}}^{LB} \leq P_{D_{xk}}^{UB} \\[2mm] 0\; \text{if}\; P_{D_{xk}}^{LB} > P_{D_{xk}}^{UB}  \end{array},
\label{eq:Dist_Phi_xk}
}
\end{equation}
\begin{equation}
\normalsize{
\begin{array}{c}
\text{\normalsize{$P_{xk}=\phi_{xk} P_{D_{xk}}^{LB}$}},\\[2mm]
\forall x\in\{ 1,\,...,\, N\}, \forall k\in\{ 1,\,...,\,\hat{N}_{x}\}.
\end{array}
\label{eq:Dist_P_xk}
}
\end{equation}
\end{subequations}
Notice that if $D2D_{xk}$ is in active mode, the transmission power is set as lower bound. This is done because the lower bound is calculated by considering the estimation of the interference between different D2D links. Thus increasing the transmission power above this level would result in higher interference between D2D links limiting the density of active D2D pairs. As a result we select the active transmission power as the lower bound to minimize the interference and maximize the density of D2D links.

In this solution the BS needs to broadcast a limited number of parameters that are common to all D2D links, thus the amount of signaling overhead introduced is significantly lower compared to a solution where CSI needs to be exchanged between the BS and each D2D link. In Appendix \ref{imp_PCSID} we give a detailed summary of the parameters that need to be calculated by the BS and D2D links to apply the algorithm.

\begin{figure}[t]
\begin{center}
\includegraphics[width=0.45\textwidth]{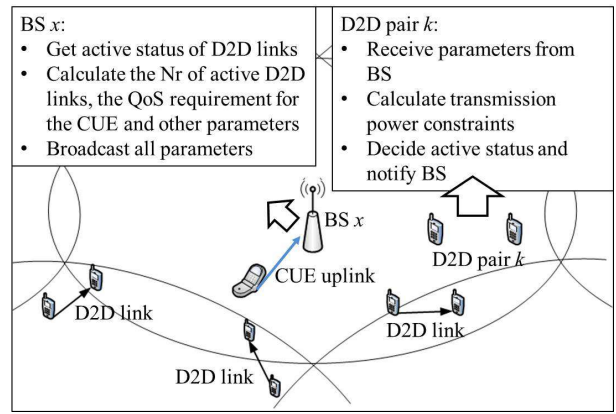}
\end{center}
\caption{Implementation of the DAC method}
\label{fig:Dist_alg}
\end{figure}

\begin{figure}[!t]
\begin{center}
\captionsetup{width=0.2\textwidth}
        \subfloat[At $D2D_{xk}$.]{%
           \label{figs:PCSI-D_imp_D2D}
		\includegraphics[width=0.235\textwidth]{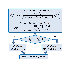}
        }%
		\subfloat[At $BS_{x}$.]{%
		   \label{figs:PCSI-D_imp_BS}
		    	\includegraphics[width=0.23\textwidth]{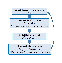}
		}%
\end{center}
\caption{Flow chart of DAC method implementation.}
\label{fig:PCSI-D_imp}
\end{figure}

To illustrate the main concept of the DAC method Fig. \ref{fig:Dist_alg} depicts the roles of a D2D pair and its serving BS. The BS keeps track of the number of active D2D links, calculates the amount of interference that the CUE uplink can tolerate (defined as $\hat{I}_{x0}^{th}$) and other necessary parameters so that the D2D links can calculate their transmission power constraints. At the same time the D2D pairs receive the parameters, calculate their transmission power constraints and notify the BS after their active status is decided.

Fig. \ref{figs:PCSI-D_imp_D2D} presents a flow chart of the specific steps of the DAC algorithm at $D2D_{xk}$. First all necessary information is collected from the BS, then $D2D_{xk}$ calculates the necessary parameters to estimate the interference conditions and the transmission power constraints. Finally $D2D_{xk}$ decides its active status and notifies the BS of its decision. Similarly Fig. \ref{figs:PCSI-D_imp_BS} depicts the necessary steps taken by $BS_{x}$ to support the DAC method. First $BS_{x}$ collects the active status of D2D links and calculates the number of active D2D links in the cell and in each sector. Then $BS_{x}$ calculates $\hat{I}_{x0}^{th}$ for the QoS of CUE and $\mathbb{E}[P_{tx_{CUE}}]$ for the expectation of the interference caused by CUEs (see Appendix \ref{imp_PCSID}), finally it broadcasts all necessary parameters or sends them in D2D discovery messages.

%



\section{Optimal Admission Control}
\label{F_CSIOp}

In this section, we present the optimal admission control assuming complete CSI is available. This serves as the performance benchmark and it is based on the same approach found in \cite{Int_const_D2D}. The goal is to maximize the frequency reuse, i.e., the number of active D2D links in the system while providing QoS to CUEs and D2D links. So a mixed integer programming (MIP) optimization problem is formulated as follows:

\begin{equation}
\normalsize{
\max_{\left(\substack{\phi_{ij}\in\left\{0,1\right\}\\ P_{ij}\in\mathbb{R}^{+}}\right)}\left\{ \sum_{i=1}^{N}\sum_{j=1}^{\hat{N_{i}}}\phi_{ij}\right\},
}
\label{eq:OP_max_phi_g}
\end{equation}
Subject to:
\begin{subequations}
\label{eq:OP_const}
\begin{equation}
\Gamma_{x0}=\,\frac{P_{x0}G_{x0x0}}{I_{x0}^{D2D}\,+\,I_{x0}^{CUE}\,+\, \mathcal{N}_{BS}}\,\geq\,\gamma_{x0}^{th},
\label{eq:OP_CUE_SINR_c}
\end{equation}
\begin{equation}
\Gamma_{xk}\,=\,\frac{\phi_{xk}P_{xk}G_{xkxk}+\left(1-\phi_{xk}\right)M_{xk}}{I_{xk}^{D2D}+I_{xk}^{CUE}+\mathcal{N}_{D}}\,\geq\,\gamma_{xk}^{th},
\label{eq:OP_D2D_SINR_c}
\end{equation}
\begin{equation}
\normalsize{\phi_{xk}P_{xk}\,\leq\,P_{D}^{max},
}
\label{eq:OP_D2D_max_P_g}
\end{equation}
\end{subequations}
\begin{center}\normalsize{$\forall x\in\{ 1,\,...,\, N\}\;\forall k\in\{ 1,\,...,\,\hat{N}_{x}\} $}.\end{center}

$M_{xk}$ is a set parameter defined in such a way that when a D2D link is inactive $\phi_{xk}=0$, its SINR constraint is always satisfied. By doing this we avoid the issue of having an unfeasible optimization problem because of inactive D2D links. The constraint for $M_{xk}$ is given by:
\begin{equation}
\normalsize{
\begin{array}{l}
M_{xk}\geq\gamma_{xk}^{th}\Big(P_{D}^{max}\left(\sum_{i=1}^{N}\sum_{j=1}^{\hat{N}_{i}}G_{ijxk} - G_{xkxk}\right) \\ \qquad\quad\;+\,P_{C}^{max}\sum_{i=1}^{N}G_{i0xk}\,+\,\mathcal{N}_{D}\Big),
\end{array}
}
\label{eq:OP_M_xk_g}
\end{equation}
where $P_{C}^{max}$ represents the maximum transmission power of CUEs. The optimization variables are the state of D2D links $\phi_{xk}$ and their transmission power $P_{xk}$.

The MIP optimization problem depicted in \eqref{eq:OP_max_phi_g} and \eqref{eq:OP_const}  cannot be solved directly  because the constraint \eqref{eq:OP_D2D_max_P_g} is nonlinear, given that $\phi_{xk}$ is a binary variable. To obtain a linear constraint we define:
\begin{equation}
\normalsize{
\tilde{P}_{xk} = \phi_{xk}P_{xk}\,\leq\,P_{D}^{max}.
}
\label{eq:OP_P_t_xk_g}
\end{equation}

Furthermore the constraints \eqref{eq:OP_CUE_SINR_c} and \eqref{eq:OP_D2D_SINR_c} are also nonlinear, thus we combine \eqref{eq:OP_P_t_xk_g} and \eqref{eq:Int_terms} to rewrite the optimization problem as:
\begin{equation}
\normalsize{
\max_{\left(\substack{\phi_{ij}\in\left\{0,1\right\}\\ \tilde{P}_{ij}\in\mathbb{R}^{+}}\right)}\left\{ \sum_{i=1}^{N}\sum_{j=1}^{\hat{N_{i}}}\phi_{ij}\right\},
}
\label{eq:OP_max_phi_l}
\end{equation}
Subject to:
\begin{subequations}
\label{eq:OP_Li_const}
\begin{equation}
\sum_{i=1}^{N}\sum_{j=1}^{\hat{N}_{i}}\tilde{P}_{ij}G_{ijx0} \leq \sum_{i=1}^{N}P_{i0}A_{ix0}^{C} - \mathcal{N}_{BS},
\label{eq:CUE_SINR_l}
\end{equation}
\begin{equation}
\sum_{i=1}^{N}\sum_{j=1}^{\hat{N}_{i}}\tilde{P}_{ij}A_{ijxk}^{D} + \phi_{xk}M_{xk} \leq B_{xk}^{D},
\label{eq:DUE_SINR_l}
\end{equation}
\begin{equation}
\normalsize{\tilde{P}_{xk}\,\leq\,P_{D}^{max},
}
\label{eq:OP_D2D_max_P_l}
\end{equation}
\end{subequations}
\begin{center}\normalsize{$\forall x\in\{ 1,\,...,\, N\}\;\forall k\in\{ 1,\,...,\,\hat{N}_{x}\}, $}\end{center}
where
\begin{subequations}
\label{eq:OP_Li_const_param}
\begin{equation}
\normalsize{
A_{ix0}^{C} = \Bigg\{\begin{array}{l}\frac{\text{\normalsize{$G_{x0x0}$}}}{\text{\normalsize{$\gamma_{x0}^{th}$}}}\;\;\, \text{if}\; i=x \\ -G_{i0x0}\; \text{if}\; i\neq x\end{array},
}
\label{eq:OP_A_C_l}
\end{equation}
\begin{equation}
\normalsize{
A_{ijxk}^{D} = \Bigg\{\begin{array}{l}\text{\normalsize{$-G_{xkxk}$}}\;\;\, \text{if}\; \left(i=x \:\text{and} \:j=k\right) \\ \text{\normalsize{$\gamma_{xk}^{th}G_{ijxk}$}}\; \text{if}\; \left(i\neq x\;\text{or} \;\;\: j\neq k\right)\end{array},
}
\label{eq:OP_A_D_l}
\end{equation}
\begin{equation}
\normalsize{
B_{xk}^{C}=M_{xk} - \gamma_{xk}^{th}\left(\sum_{i=1}^{N}P_{i0}G_{i0xk} + \mathcal{N}_{D}  \right).
}
\label{eq:OP_B_D_l}
\end{equation}
\end{subequations}
The optimization problem found in \eqref{eq:OP_max_phi_l}, \eqref{eq:OP_Li_const} and \eqref{eq:OP_Li_const_param} is a mixed integer linear programming (MILP) problem and its solution can be found by using exhaustive search algorithms.



\section{Numerical analysis}
\label{Num_results}

%
%

 To evaluate the performance of the proposed interference coordination methods we conduct extensive Monte-Carlo simulations. We consider 7 circular cells of radius $R$ where BSs are located at the center of the cells. In order not to underestimate the interference conditions we only collect data from the center cell. In each realization we generate one CUE uniformly distributed per cell, where the distance to the BS is within $d_{CUE} \in [d_{min}, R]$. Also $\hat{N}$ D2D pairs are generated following an uniform distribution where the distance between devices of the same pair is randomly selected within $d_{D2D} \in [D_{min}, D_{max}]$. The channel model accounts for path loss and log-normal shadow fading implemented according to 3GPP specifications \cite{3GPP_channel}. Notice that our analysis is not sensitive to the specific density of users in the system. The reason for this is that first, the density of D2D nodes is not of high importance given that most of them may be silent, what it is important is the maximum density of D2D nodes that can be active at the same time in the network, which is determined by the admission control algorithms. Second, since we have included log-normal shadow fading in our channel model, the received powers and interference terms will not be tightly dependent on the user distribution given that the log-normal distribution has quite long tails. Table \ref{tab:sim_param} summarizes the main parameters used in the simulations.
\begin{table}[!t]
\caption{Simulation parameters.}
\label{tab:sim_param}

\small{
\begin{center}
\begin{tabular}{m{3.75cm} m{4.1cm}}
\textbf{Description} & \center{\textbf{Representation and Value}}\tabularnewline
\hline
Radius
&
\center{$R=$400 [m]}
\tabularnewline
Noise power
&
\center{$\mathcal{N}_{D},\mathcal{N}_{BS}=-$174[dBm/Hz]}
\tabularnewline
RB bandwidth
&
\center{$B_{w}=$180 [kHz]}
\tabularnewline
Carrier frequency
&
\center{$f_{c}=$2 [GHz]}
\tabularnewline
Max. transmission power
&
\center{$P_{D}^{max},P_{C}^{max}=$23 [dBm]}
\tabularnewline
Min. distance between the BS and the users.
&
\center{$d_{min} = $10 [m]}
\tabularnewline
D2D distance ($d_{D2D}$) bounds
&
\center{$D_{min}=$ 10 [m]\\[2pt]$D_{max}= $ 40 [m]}
\tabularnewline
Number of cells
&
\center{$N = $ 7}
\tabularnewline
Available D2D pairs per cell
&
\center{$\hat{N} = $ 10}
\tabularnewline
Path loss coeff. (user to BS)
&
\center{$c_{0}=-$30.55 [dB]}
\tabularnewline
Path loss coeff. (user to user)
&
\center{$c_{d}=-$28.03 [dB]}
\tabularnewline
Path loss exp. (user to BS)
&
\center{$\alpha_{0}=$ 3.67}
\tabularnewline
Path loss exp. (user to user)
&
\center{$\alpha_{d}=$ 4}
\tabularnewline
Monte-Carlo realizations
&
\center{$5000$}
\tabularnewline
\hline
\end{tabular}
\end{center}
}
\end{table}
 The MILP optimization problem is solved by using the optimization software MOSEK implemented in MATLAB.

 To present a comparative analysis with previous solutions we introduce two single-cell methods provided in \cite{Int_const_D2D}. The first is an optimization problem formulated with full CSI within each cell and the second is a centralized solution where a statistical upper bound  for the number of active D2D links is derived assuming no available CSI. The initial solution is denoted as peak interference constraint (PIC) in a single-cell ``PIC. Single-cell" and the latter as average interference constraint (AIC) in a single-cell ``AIC. Single-cell". These two solutions were developed for a single-cell scenario, thus we implement them independently in each cell.

 The transmission power of CUEs is given by the LTE OFPC \cite{LTE_PC_VTC} depicted as:
  \begin{equation}
  \begin{array}{c}
  P_{x0_\text{dBm}}\,=\,P_{0_\text{dBm}}-\alpha_{p}G_{x0x0_\text{dB}},\\[2mm]
  P_{0_\text{dBm}} = \alpha_{p}\left(\gamma_{CUE_\text{dB}}^{th}+\mathcal{N}_{BS_\text{dB}}\right)+\left(1-\alpha_{p}\right)\left(P_{C_{dBm}}^{max} \right),
  \end{array}
  \label{eq:P_CUE}
  \end{equation}
  where $\alpha_{p}$ is the path loss compensation factor and $\gamma_{CUE_\text{dB}}^{th}$ is the open loop target signal-to-noise ratio (SNR). We assume that the transmission powers of CUEs are given and cannot be modified by the proposed interference coordination methods.

\begin{figure}[!h]
       \begin{center}
        \captionsetup{width=0.2\textwidth}
          \subfloat[Max. Nr. D2D links vs Target CUE SINR Loss $\delta$ for $d_{D2D_{max}}=40$  \text{[m]}.]{%
             \label{fig:sfig:NUB_delta}
             \includegraphics[width=0.23\textwidth]{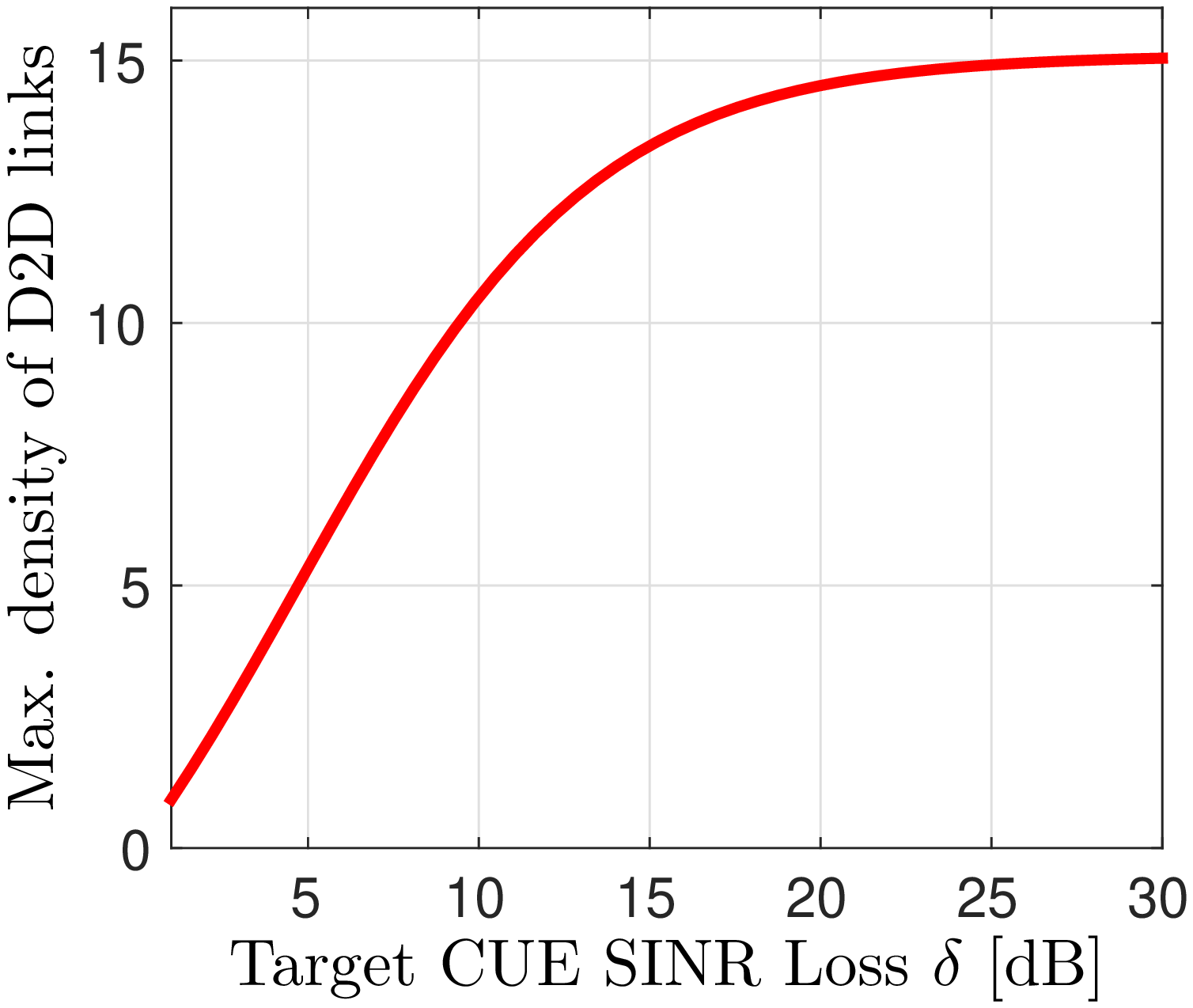}
          }
          \subfloat[Upper bound Nr. D2D links vs D2D max. distance $d_{D2D_{max}}$  for $\delta = 3$ \text{[dB]}.]{%
             \label{fig:sfig:NUB_d_D2D}
             \includegraphics[width=0.23\textwidth]{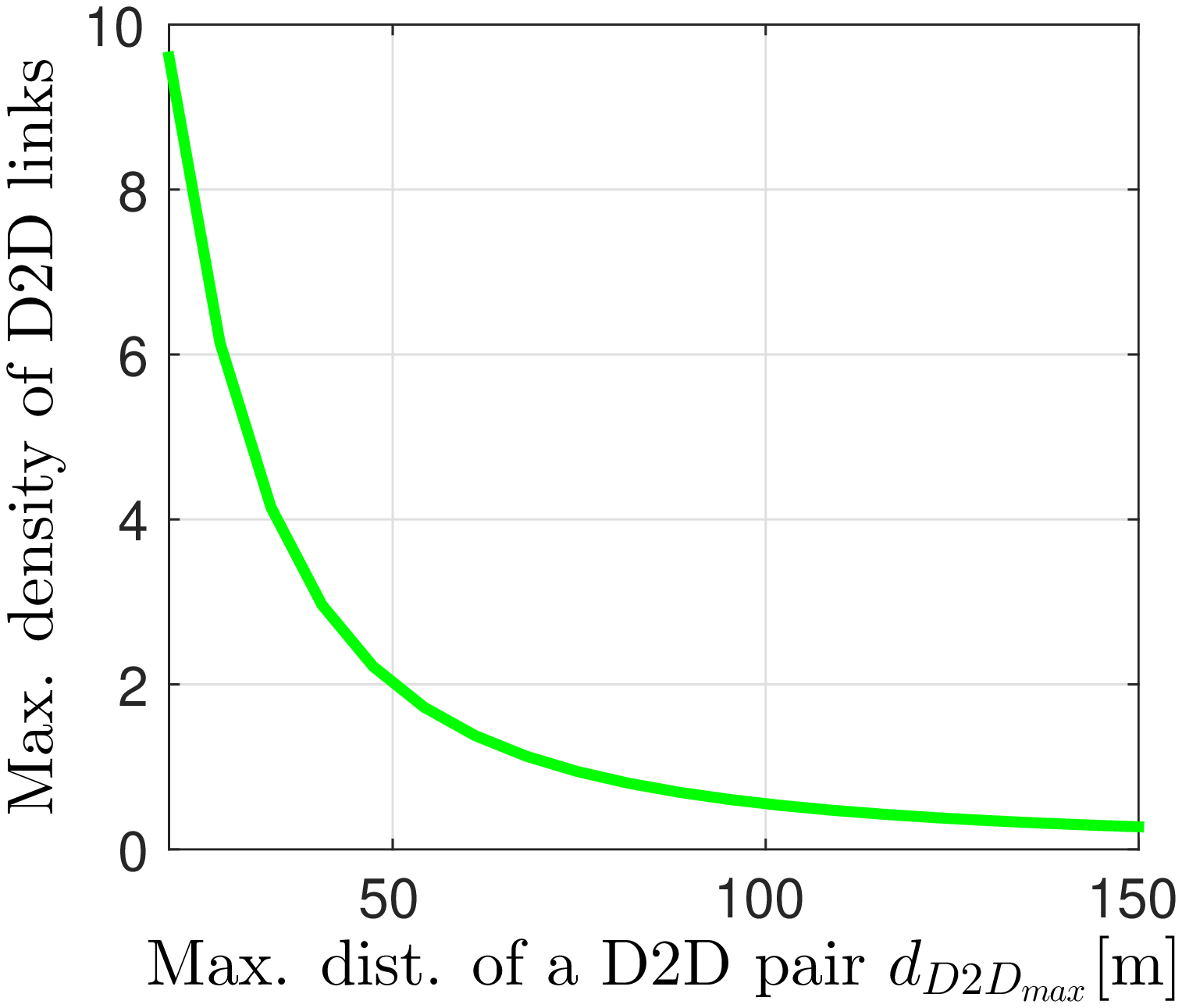}
          }%

      \end{center}
      \caption{Upper Bound of number of D2D links, for $\gamma_{D} = 8$ \text{[dB]}, in a circular area of radius $R$.}%
    \label{fig:NUB_theo}
  \end{figure}

   Fig. \ref{fig:NUB_theo} depicts the behavior of the theoretical upper bound for the number of D2D links, in a circular area if radius R, derived in \eqref{eq:N_UB_a} versus the target CUE SINR loss and maximum distance between devices within the same D2D pair. In Fig. \ref{fig:sfig:NUB_delta} as $\delta$ increases the number of D2D links also increases until it reaches a point of saturation. This occurs because as $\delta$ increases the constraint regarding the QoS of CUEs becomes less strict and more D2D links can be admitted in the system. However after a certain point no more D2D links can be admitted due to the interference caused by D2D links towards each other.  Fig. \ref{fig:sfig:NUB_d_D2D} shows that as the distance $d_{D2D_{max}}$ becomes higher the number of D2D links decreases exponentially, thus the most gain of having D2D links is obtained when the D2D transmitter and receiver of the same pair are in proximity to each other. This result shows that the distance between the transmitter and receiver of the same D2D pair and the tolerable performance loss of CUEs play a key role in the maximum number of D2D links that the system can support.

Now we compare the overall performance of the different interference coordination methods for a low target CUE SINR loss ($\delta =2$ [dB]) and a high D2D SINR target ($\gamma_{D} = 16$[dB]). These conditions represent the ideal case of D2D communications where the D2D links achieve high data rates while causing the least amount of disturbance towards the QoS of CUEs.

\begin{figure}[t]
     \begin{center}
\captionsetup{width=0.2\textwidth}
        \subfloat[CDF of CUE SINR loss]{%
           \label{fig:sfig:CDF_CUE_loss_PC}
		\includegraphics[width=0.23\textwidth]{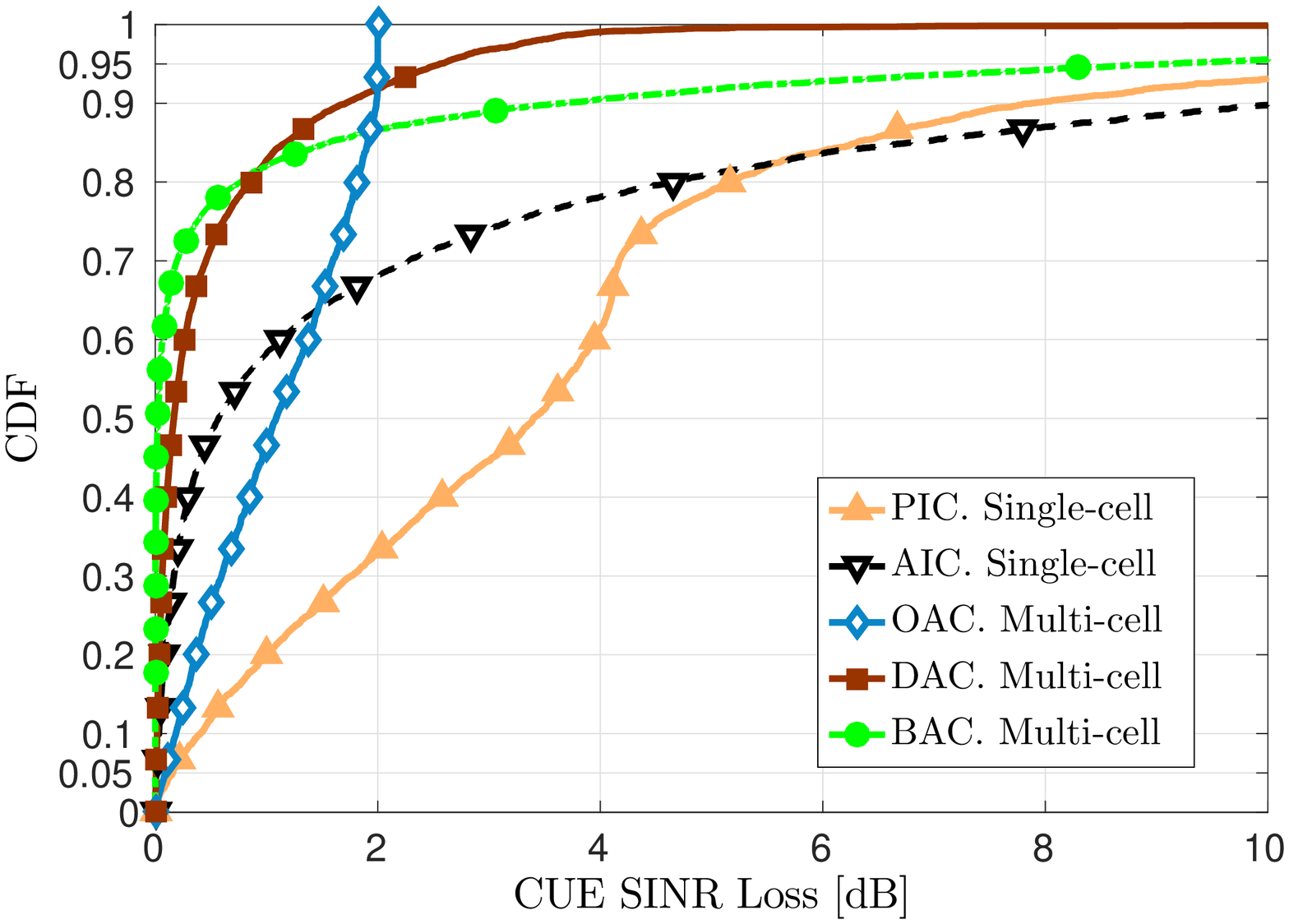}
        }%
        \subfloat[CDF of D2D links SINR.]{%
           \label{fig:sfig:CDF_D2D_SINR_PC}
	  \includegraphics[width=0.23\textwidth]{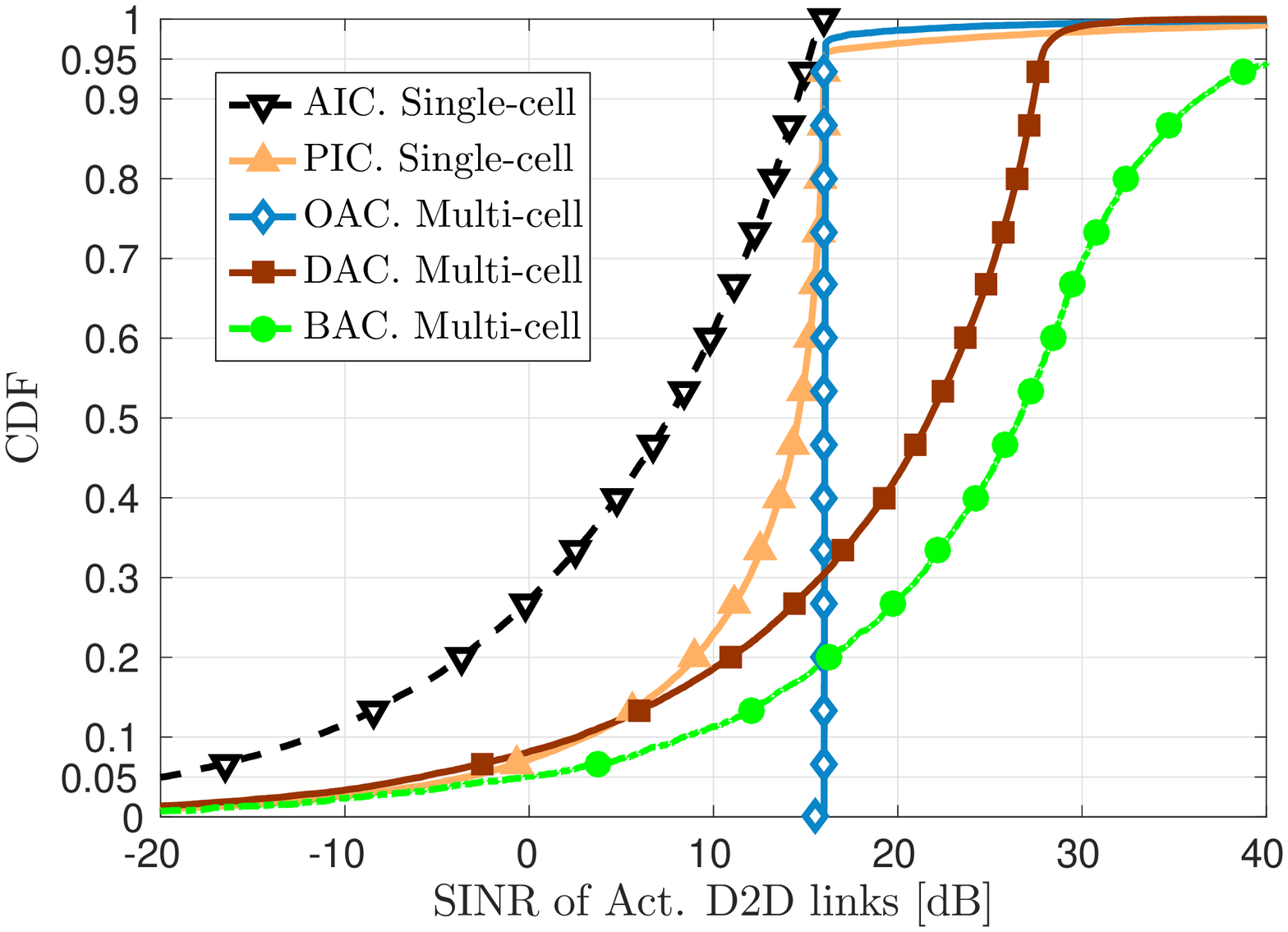}
        }\\
        \subfloat[CDF of Nr of D2D links with QoS.]{%
           \label{fig:sfig:CDF_Nr_D2DQoS_PC}
		\includegraphics[width=0.23\textwidth]{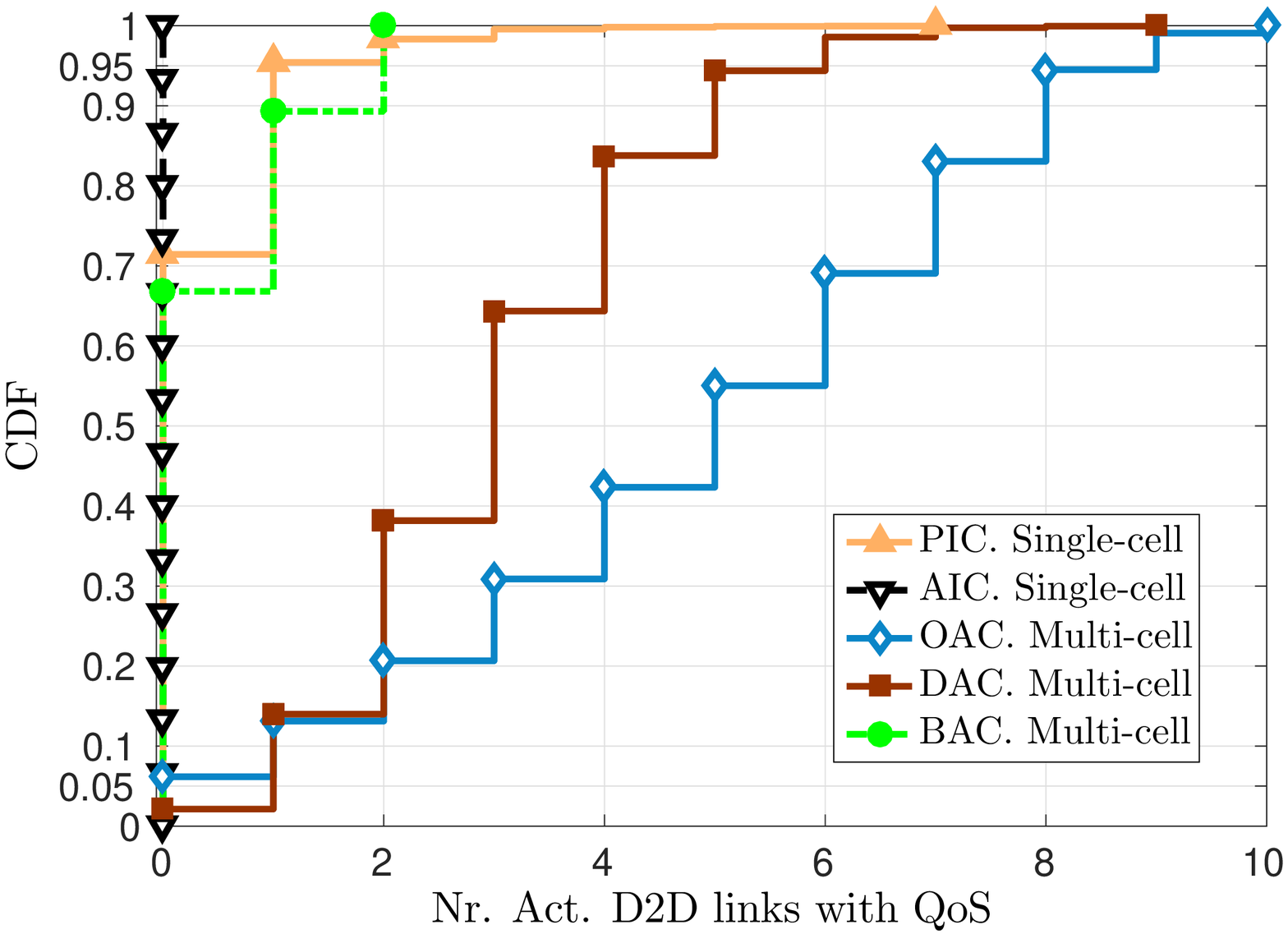}
        }%
        \subfloat[CDF of SE.]{%
           \label{fig:sfig:CDF_SE_PC}
	  \includegraphics[width=0.23\textwidth]{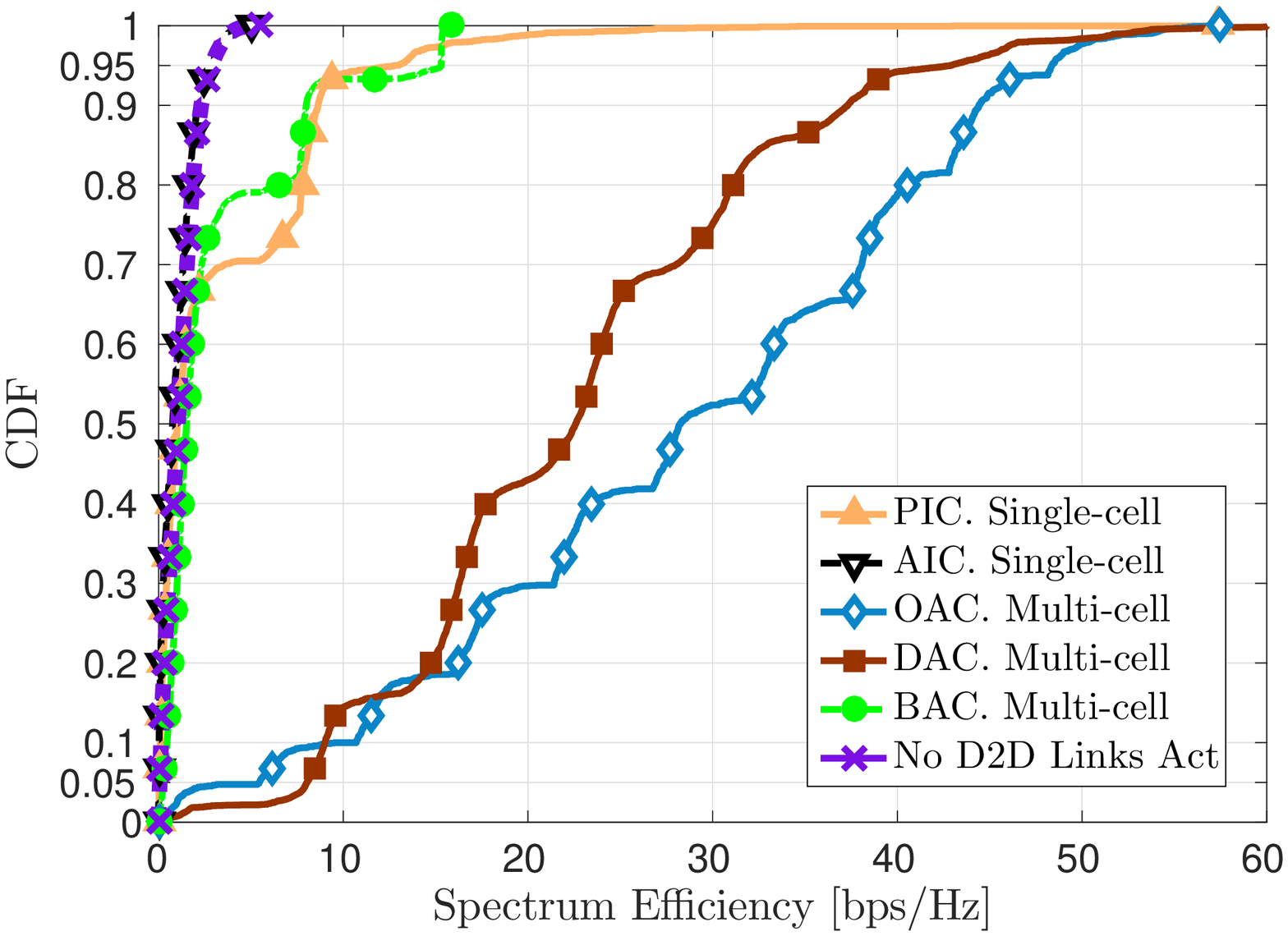}
        }%

    \end{center}
    \caption{CDF of overall performance for $\gamma_{D}=$ 16 [dB] $\delta=$ 2 [dB].}%
  \label{fig:CDF_all}
\end{figure}

 Fig. \ref{fig:sfig:CDF_CUE_loss_PC} and Fig. \ref{fig:sfig:CDF_D2D_SINR_PC} show the empirical cumulative distribution function (CDF) of the SINR loss of CUEs and the SINR of active D2D links respectively. For the single-cell  methods, the SINR of D2D links is mostly below the required target $\gamma_{D}$ and there is a significant CUE SINR loss when compared to the multi-cell solutions. Fig. \ref{fig:sfig:CDF_Nr_D2DQoS_PC} depicts the empirical CDF of the number of active D2D links with QoS and Fig. \ref{fig:sfig:CDF_SE_PC} shows the empirical CDF of the SE of the system. We see that the single-cell solutions neglect the impact of inter-cell interference when coordinating the admission and power control of D2D links, thus perform poorly.

For the BAC method we can see that the SINR of D2D links is quite high and for over 80\% of devices the required target is fulfilled, furthermore the CUE SINR loss is well below the required target for 80\% of the CUEs. However the number of D2D links with QoS and the SE are quite limited. This occurs because the BAC method does not control the individual admission of each D2D link, thus when their target SINR $\gamma_{D}$ is high the number of D2D links that can be active is low limiting the scalability and SE.

In the case of the DAC method we can see that the SINR of D2D links is more balanced around the required target with over 70\% of devices above it and the CUE SINR loss is quite low, assuring the respective target for over 90\% of the CUEs. Furthermore we can see a significant improvement in the number of D2D links combined with high values of SE.

Let the outage probability be the probability of not achieving the target SINR
\begin{equation}
P_{out_{xk}}=Prob\left\{ \Gamma_{xk} < \gamma_{xk}^{th}\right\}.
\label{eq:Pr_out_D2D}
\end{equation}
For the results shown in Fig. \ref{fig:CDF_all} we see that the outage probability assured by each of the evaluated methods is different, thus it would be interesting to compare the performance when the QoS offered by all methods is the same. To achieve this we calculate the 5 percentile of the SINR of D2D links which gives us the different values of SINR for which all methods provide a 5\% D2D outage probability. Similarly we consider the 95 percentile of the CUE SINR loss that shows the different values of CUE SINR losses for which a 5\% CUE outage probability is assured by all methods.
\begin{figure}[t]
\begin{center}
\includegraphics[width=0.45\textwidth]{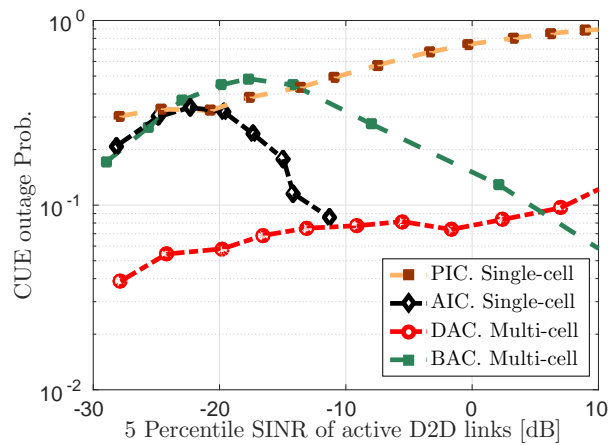}
\end{center}
\caption{CUE outage probability vs 5 \% D2D links SINR, for $\delta=$ 2 [dB].}
\label{fig:QoS_CUE}
\end{figure}

Fig. \ref{fig:QoS_CUE} depicts the CUE outage probability for different values of the 5 percentile of the SINR of D2D links to illustrate the effect of the QoS of D2D links on the QoS of CUEs. We can see that the BAC method has a concave shape behavior as the SINR of D2D links increases. This is caused by the selection of the target received power of the D2D links ($\hat{P}_{r_{D}}$) done in the BAC method. Initially as the SINR of D2D links increases more interference is caused to the CUEs increasing their outage, then for higher D2D SINR values less D2D links are allowed to be active which in turn reduces the outage probability of CUEs. For the DAC method we see that for a wide range of D2D SINR values the outage probability of CUEs is maintained at acceptable levels.

   At this point we have seen the performance of the proposed methods in terms of the QoS that they can provide to D2D links and CUEs. Now we would like to get a more comprehensive view of the performance of the system when a specific QoS is assured by all methods for all users. For this we present 3D graphs where on one axis we place the 5 percentile of the SINR of D2D links and on the other axis we put the 95 percentile of the CUE SINR loss. By making this selection of axes we can evaluate the performance for a 5\% outage probability of both CUEs and D2D links. Notice that the OAC method is not shown in the figures as it does not have outage probability since it corresponds to an optimal solution where all QoS requirements are assured for all active users.

\begin{figure}[t]
\begin{center}
\hspace*{-8mm}
\includegraphics[width=0.45\textwidth]{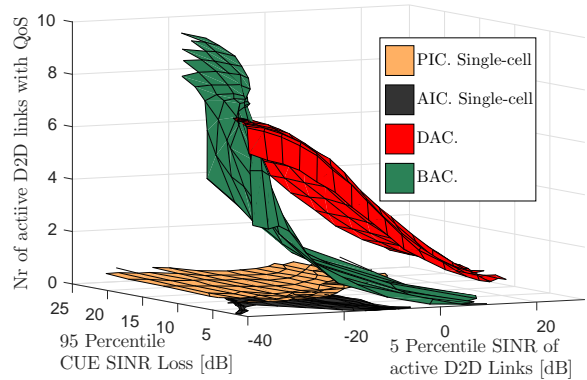}
\end{center}
\caption{Nr of active D2D links with QoS vs 5\% SINR of active D2D links and 95\% CUE SINR loss.}
\label{fig:3D_Nr_D2D}
\end{figure}

\begin{figure}[t]
\begin{center}
\hspace*{-8mm}
\includegraphics[width=0.45\textwidth]{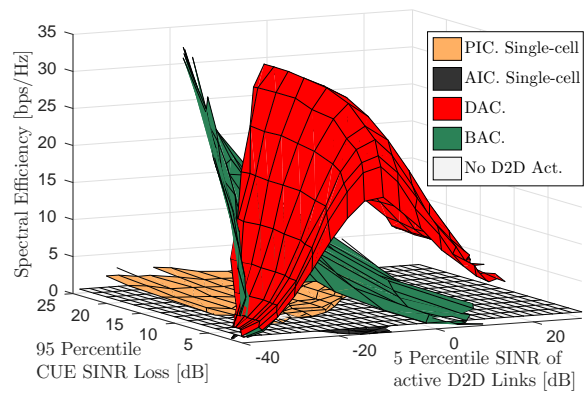}
\end{center}
   \caption{Spectral efficiency [bps/Hz] vs 5\% SINR of active D2D links and 95\% CUE SINR loss.}
\label{fig:3D_sum_rate}
\end{figure}

Fig. \ref{fig:3D_Nr_D2D} shows the performance of the number of D2D links with QoS and Fig. \ref{fig:3D_sum_rate} depicts the SE for all the evaluated methods. We see that for the DAC method as the SINR of D2D links grows, the number of active pairs with QoS decreases, which causes a quasi-concave behavior for the SE. When the SINR of D2D links is low there are more D2D links active in the system but their data rates are low as well, then as the SINR of D2D links increases the data rates also increase, resulting in higher SE. However as the SINR of D2D links increases the number of D2D links decays which causes the overall SE to drop regardless of the SINR values of D2D links. As the CUE SINR loss increases more D2D links can be admitted in the system which in turn increments the SE, however after a certain point the number of D2D links reaches a saturation point along with the SE. This means that the interference between different D2D links is becoming a more and more important limiting factor in the density of active D2D links.

Compared to the other methods, the DAC method provides a significant improvement in D2D density and SE for a wide range of D2D SINR with a low CUE SINR loss. Thus the DAC method is a low complexity solution suitable for practical D2D communications.

\section{Conclusion}
\label{Conc}

 We have conducted a comprehensive study on the admission control, interference coordination, and power control for scalable D2D communications underlay cellular networks. While there have been sophisticated ICIC techniques to handle the first level interference among cellular users, our study shows that it is of paramount importance to develop new interference coordination techniques to solve the second level interference issues, i.e., those from D2D to cellular communications and those among D2D links themselves. We have developed three interference coordination methods, i.e. OAC, DAC and BAC to maximize the network frequency reuse, while assuring QoS to all users.

The results show that the best performance is achieved by the OAC followed by the DAC and finally the BAC method. Thus we have seen that as the CSI becomes available the performance of the interference coordination is increased, however the complexity and signaling overhead is also increased. Thus, a trade-off must be found in order to provide a practical solution that can be deployed in real systems.

The DAC method has been proven to be a good candidate for practical applications. This method provides good SE and scalability for a wide range of  QoS requirements for both D2D links and CUEs and can be easily implemented in 5G LTE-A systems. We can conclude that proper design of D2D communications can dramatically increase the network frequency reuse and thus network capacity, especially for short distance communications. In addition, it is possible to achieve this with low complexity interference coordination and low power transmission methods without affecting existing cellular communications.


\appendix[Parameters of the DAC method]
\label{imp_PCSID}

%

Tables \ref{tab:DAC_param_BS} depicts all additional parameters that are calculated at the BS and broadcasted to the D2D links to support the DAC method. Similarly table \ref{tab:DAC_param_D2D}  illustrates all additional parameters that the D2D links calculate in order to obtain their transmission power constraints.


\begin{table}[!t]
\vspace*{-7mm}
\caption{Parameters broadcasted by the BS.}
\label{tab:DAC_param_BS}
\small{
\begin{center}
\begin{tabular}{|m{0.25\textwidth}| m{0.2\textwidth}|} 
\hline
\textbf{Parameter} & \textbf{Description}\\ 
\hline
\hspace*{-12pt}
\begin{tabular}{m{0.3\textwidth}}
$\mathbb{E}[I_{x0}^{CUE}]=$\\[4pt]$\frac{A_{x0}^{C}}{A_{cl_{x}}}\mathbb{E}[P_{tx_{CUE}}]\mathbb{E}[G_{CUE-BS-out}]$
\end{tabular} & Expected value of inter-cell interference caused by CUEs towards $BS_{x}$.\\[16pt]
\hline
\hspace*{-12pt}
\begin{tabular}{m{0.3\textwidth}}
$\mathbb{E}[P_{tx_{CUE}}]=$\\[2pt]$\mathbb{E}\!\left[\!\left(\!\frac{\gamma_{CUE}^{th}\mathcal{N}_{BS}}{G_{CUE-BS-in}}\!\right)^{\!\!\alpha_{p}}\!\!(P_{C}^{max})^{(1-\alpha_{p})}\!\right]$
\end{tabular} & Expected value of the transmission power of CUEs.\\[16pt]
\hline
\hspace*{-12pt}
\begin{tabular}{m{0.3\textwidth}}
$\mathbb{E}[G_{CUE-BS-in}]=$\\[4pt]$\frac{2c_{0}\left(d_{CUE-BS_{min}}^{-(\alpha_{0}-2)}-R^{-(\alpha_{0}-2)}\right)}{R^{2} \left(\alpha_{0}-2\right)}$
\end{tabular}  & Expected value of the channel gain between a CUE and its serving BS.\\[14pt]
\hline
\hspace*{-12pt}
\begin{tabular}{m{0.3\textwidth}}
$\mathbb{E}[G_{CUE-BS-out}]=$\\[4pt]$\frac{2c_{0}\left(R^{-(\alpha_{0}-2)}-d_{CUE-BS_{max}}^{-(\alpha_{0}-2)}\right)}{(d_{D2D-BS_{max}})^{2} \left(\alpha_{0}-2\right)}$
\end{tabular} & Expected values of the channel gain between an interfering CUE and a given BS.\\[14pt]
\hline
\hspace*{-4pt}
$d_{CUE-BS_{max}} = \left(\frac{P_{C}^{max}c_{0}}{\mathcal{N}_{BS}}\right)^{1/\alpha_{0}}$ & Max. distance between an interfering CUE and a given BS.\\[8pt]
\hline
\hspace*{-4pt}
$ A_{x0}^{C} = \pi (d_{CUE-BS_{max}})^2 - \pi R^2$ & Interference area for $BS_{x}$ as the receiver, when CUEs are the source of interference.\\
\hline
\end{tabular}
\end{center}
}
\vspace*{-7mm}
\end{table}
\begin{table}[!t]
\vspace*{-7mm}
\caption{Parameters calculated by the D2D links.}
\label{tab:DAC_param_D2D}
\small{
\begin{center}
\begin{tabular}{|m{0.23\textwidth}| m{0.22\textwidth}|} 
\hline
\hspace*{-10pt}
\begin{tabular}{m{0.3\textwidth}}
$\mathbb{E}[I_{xk}^{CUE}]=$\\[4pt]$\frac{A_{xk}^{C}}{A_{cl_{x}}}\, \, \mathbb{E}[P_{tx_{CUE}}]\,\mathbb{E}[G_{CUE-D}]$
\end{tabular} & Expected value of interference caused by CUEs towards $D2D_{xk}$.\\[16pt]
\hline
\hspace*{-10pt}
\begin{tabular}{m{0.3\textwidth}}
$\mathbb{E}[G_{CUE-D}]=$\\[4pt]$\frac{2c_{d}\left(d_{CUE-D_{min}}^{-(\alpha_{d}-2)}\!-d_{CUE-D_{max}}^{-(\alpha_{d}-2)}\right)}{(d_{CUE-D_{max}})^{2} \left(\alpha_{d}-2\right)}$
\end{tabular}  & Expected value of the channel gain between a given D2D link and a CUE.\\[14pt]
\hline
\hspace*{-2pt}

$d_{CUE-D_{max}}\!=\!\left(\frac{P_{C}^{max}c_{d}}{\mathcal{N}_{D}}\right)^{\frac{1}{\alpha_{d}}}$ & Max. distance between a given CUE and a D2D link.\\[8pt]
\hline
\hspace*{-2pt}
$A_{xk}^{C} = \pi (d_{CUE-D_{max}})^2$ & Interference area for $D2D_{xk}$ as a receiver, when CUEs are the source of interference.\\
\hline
\hspace*{-2pt}
$A_{x0} = \pi (d_{D2D-BS_{max}})^2$ & Interference area for $BS_{x}$ as the receiver, when D2D links are the source of interference.\\
\hline
\hspace*{-2pt}
$d_{D2D-BS_{max}}\!=\!\left(\!\frac{P_{D}^{max}c_{0}}{\mathcal{N}_{BS}}\!\right)^{\frac{1}{\alpha_{0}}}$ & Max. distance between an interfering D2D link and a given BS.\\[8pt]
\hline
\hspace*{-2pt}
$A_{xk} = \pi (d_{D2D-I_{max}})^2$ &  Interference area for $D2D_{xk}$ as the receiver, when D2D links are the source of interference.\\
\hline
\hspace*{-2pt}
$d_{D2D-I_{max}}=\left(\frac{P_{D}^{max}c_{d}}{\mathcal{N}_{D}}\right)^{\frac{1}{\alpha_{d}}}$ &  Max. distance between an interfering D2D link and another given D2D pair.\\
\hline
\end{tabular}
\end{center}
}
\vspace*{-7mm}
\end{table}
\end{document}